\documentclass[switch, nocrop]{article} % Method A for two-column formatting
\usepackage{preprint}

\usepackage{etoolbox}
\AtBeginDocument{%
  \setlength{\headsep}{20pt}
}

\usepackage[T1]{fontenc}
\usepackage{lmodern}
\usepackage{graphicx}      % include this line if your document contains figures
\usepackage{textgreek}
\usepackage{amssymb}
\usepackage{amsmath}
\usepackage[ruled,vlined,algo2e]{algorithm2e}
\usepackage{xcolor}
\usepackage{import}
\usepackage{mathrsfs}
\usepackage{bm}
\usepackage{multirow}
\usepackage{placeins}
\usepackage{afterpage}
\usepackage{hyperref}
\usepackage{orcidlink}

\usepackage{tabularx} % in the preamble
\usepackage[flushleft]{threeparttable}
\usepackage{mdframed}
\usepackage{multicol}
\usepackage{nomencl}
\usepackage{framed}
\usepackage[most]{tcolorbox}
\usepackage{lipsum}

\usepackage{subcaption}
\usepackage{cleveref}

\usepackage{layouts}
\usepackage{paralist}

\usepackage{xfrac}

\usepackage[flushleft]{threeparttable}

\usepackage[width=\textwidth]{caption}

\renewenvironment{itemize}[1]{\begin{compactitem}#1}{\end{compactitem}}

\renewcommand\nomgroup[1]{%
	\item[\bfseries
	\ifstrequal{#1}{A}{Greek letters}{%
	\ifstrequal{#1}{B}{Abbreviations}{}}%
]}
\renewcommand{\nompreamble}{\begin{multicols}{2}}
\renewcommand{\nompostamble}{\end{multicols}}
\makenomenclature

\DeclareMathOperator*{\minA}{min}

\newcommand{\CommentMmo}[1]{\textcolor{green}{ #1 }}
\renewcommand{\CommentMmo}[1]{}

\title{Efficient Adjoint Petrov-Galerkin Reduced Order Models for fluid flows governed by the incompressible Navier-Stokes equations}

\usepackage{authblk}

\author[1\thanks{\tt{Kamil.Sommer@rub.de}}]{Kamil David Sommer\orcidlink{0000-0003-3594-2981}}

\author[1]{Lucas Mieg\orcidlink{0000-0002-3516-6890}}

\author[2]{Siddharth Sharma\orcidlink{0009-0001-4552-3257}}

\author[2]{Romuald Skoda}

\author[1]{Martin M\"onnigmann\orcidlink{0000-0002-9277-2061}}

\affil[1]{Automatic Control and Systems Theory, Ruhr-Universit\"at Bochum, Bochum, Germany}
\affil[2]{Hydraulic Fluid Machinery, Ruhr-Universit\"at Bochum, Bochum, Germany}

\begin{document}

\maketitle

\begin{abstract}
	This research paper investigates the Adjoint Petrov-Galerkin (APG) method for reduced order models (ROM) and fluid dynamics governed by the incompressible Navier-Stokes equations. The Adjoint Petrov-Galerkin ROM, derived using the Mori-Zwanzig formalism, demonstrates superior accuracy and stability compared to standard Galerkin ROMs. However, challenges arise due to the time invariance of the test basis vectors, resulting in high computational requirements. To address this, we introduce a new efficient Adjoint Petrov-Galerkin (eAPG) ROM formulation, extending its application to the incompressible Navier-Stokes equations by exploiting the polynomial structure inherent in these equations. The offline and online phases partition eliminates the need for repeated test basis vector evaluations. This improves computational efficiency in comparison to the general Adjoint Petrov-Galerkin ROM formulation. A novel approach to augmenting the memory length, a critical factor influencing the stability and accuracy of the APG-ROM, is introduced, employing a data-driven optimization. Numerical results for the 3D turbulent flow around a circular cylinder demonstrate the efficacy of the proposed approach. Error measures and computational cost evaluations, considering metrics such as floating point operations and simulation time, provide a comprehensive analysis.
\end{abstract}
 % (optional)
\vspace{0.35cm}

%###########################

\section{Introduction}\label{sec:Introduction}
In modern engineering, the important role played by high-fidelity numerical simulations, and especially those addressing fluid flow dynamics governed by the Navier-Stokes equations, cannot be overrated. However, the excessive computational costs associated with full order models (FOM) present a significant hurdle for many applications. Recognizing the impracticality of extensive computational requirements, there has been a notable shift towards the development and use of reduced order models (ROM). 
The primary objective lies in balancing computational efficiency with accuracy, a trade-off that is central to the implementation of ROMs. An integral aspect of reduced order modeling involves defining reduced spaces in which computational cost and accuracy can be systematically controlled.
Among the numerous techniques used for reduced order modeling, projection-based ROMs using the Proper Orthogonal Decomposition (POD) and Galerkin Projection (GP) are popular, particularly in the realm of fluid dynamics, although not exclusively limited to it. 

POD is instrumental in constructing a set of orthonormal basis vectors from information extracted from high-fidelity simulations. These basis vectors form the foundation for projection-based ROMs. The order reduction is achieved through the truncation of a substantial number of seemingly insignificant basis vectors, a process designed to retain only a select few basis vectors. We call the basis vectors that are and are not retained the resolved coarse-scale and unresolved fine-scale basis vectors, respectively. The truncation ensures that the subsequent Galerkin Projection yields a Galerkin reduced order model (G-ROM) with significantly fewer degrees of freedom than the FOM. Certainly, the GP stands out as one of the most straightforward approaches in reduced order modeling. Notably, it simplifies the modeling process by employing the same set of basis vectors for both approximation and projection. This significantly mitigates the computational complexity associated with the numerical simulation, making it more applicable to practical applications that are subject to real-world constraints such as real-time processing or limited computational resources. Examples include contaminant airflow~\cite{John2010}, magnetic resonance imaging~\cite{Seoane2020}, the flow around circular and oscillating cylinders and grooved channels~\cite{Bergmann2008, Deane1991, Liberge2010}, the flow inside of centrifugal and positive replacement pumps~\cite{Gunder2018, Sommer2023}, diffusion and heat conduction problems in drying processes~\cite{Berner2017, Berner2020}, and thermal flows in integrated circuits~\cite{Meyer2017}, to name only a few.

However, the apparent simplicity of G-ROMs can come at a cost. Such models often struggle with stability and accuracy issues, particularly when applied to highly turbulent flows (see, e.g.,~\cite{Akhtar2012, Baiges2015}). The lack of stability and accuracy in simple G-ROMs can be attributed to a key factor - the drastic truncation of basis vectors (see, e.g.,~\cite{Ahmed2021, Couplet2003, Bergmann2009}). This truncation gives rise to the ROM closure problem that can be interpreted as a model error, related to the challenges encountered in simulations of turbulent flows where closure modeling is required, as seen in Large Eddy Simulations (see, e.g.,~\cite{Sagaut2006}). The analogy with turbulent flows is evident, as both scenarios involve operating in a coarse, under-resolved framework where closure modeling becomes crucial. In the case of ROMs, the indirect connections arise from the nonlinearity of dynamic systems, where the truncated modes in fact interact with the retained modes. A drastic modal truncation for computational efficiency unintentionally suppresses these interactions (see, e.g.,~\cite{Akhtar2012, Baiges2015}).

To address the closure problem in ROMs, various approaches have been proposed, including mixing-length ROM closure (see, e.g.,~\cite{Aubry1988, Couplet2003}), Smagorinsky ROM closure (see, e.g.,~\cite{Ullmann2010}), spatial filtering (see, e.g.,~\cite{Xie2016, Xie2018, Wells2017, Sabetghadam2012}), stochastic closure (see, e.g.,~\cite{Themistoklis2009, Resseguier2015}), data-driven least-squares regression (see, e.g.,~\cite{Couplet2005, Buffoni2006, Perret2006, Zucatti2021}) and non-intrusive neural network regression (see, e.g.,~\cite{San2015, Pawar2019, Mjalled2023a, Mjalled2023b}), and multiscale approaches (see, e.g.,~\cite{Borggaard2008, Wang2012b, Stabile2019, Reyes2020, Mou2021}). Each method offers a unique perspective and set of techniques to mitigate the challenges posed by the closure problem, aiming to strike a balance between computational efficiency and the accurate representation of complex system dynamics.

A particularly noteworthy approach is the Petrov-Galerkin Projection which involves the application of two distinct sets of basis vectors. Although the literature on Galerkin-based ROMs is extensive, comparatively fewer ROMs have been developed using Petrov-Galerkin projection, including the least squares Petrov-Galerkin ROM~\cite{Carlberg2011, Carlberg2017}, streamline upwind Petrov-Galerkin ROM~\cite{Kragel2005, Bergmann2009, Giere2015}, Petrov-Galerkin ROM with nonlinear optimization based on manifolds~\cite{Reineking2022}, and the Adjoint Petrov-Galerkin ROM~\cite{Parish2020}.

Parish et al.~\cite{Parish2020} deals with the derivation of a Petrov-Galerkin Projection approach using the Mori-Zwanzig formalism. The authors construct a ROM exclusively with the resolved coarse-scale basis vectors and introduce a memory term to model the influence of truncated unresolved fine scales. This memory term depends on the temporal history of the resolved coarse-scale variables. Parish et al. establish a ROM known as the Adjoint Petrov-Galerkin ROM (APG-ROM), which is based on a new set of test basis vectors. It has been demonstrated that the accuracy and stability of the APG-ROM considerably surpass those achieved with a standard G-ROM~\cite{Parish2020}. The main drawback of the APG-ROM is the time invariance of the introduced test basis vectors, which implies that frequent additional calculations in the spatial domain must be carried out. The dimension of the ROM is several orders of magnitude smaller than that of the FOM. Consequently, while the ROM is designed to reduce computational effort, the need for frequent calculations in the spatial domain undermines this benefit, particularly when dealing with a high number of discretized spatial locations. 

This paper aims to improve the APG-ROM formulation and extends its applicability to the incompressible Navier-Stokes equations. This is achieved by partitioning the computations into distinct offline and online phases. This division is a common practice for standard G-ROMs of fluid flows. By leveraging the polynomial nature of the Navier-Stokes equations, the computationally demanding operations of the APG-ROM in the spatial dimension can be executed during the offline phase. This eliminates the need for the repeated evaluation of test basis vectors in each time step of the numerical integration in the online phase. We call our approach efficient APG-ROM (eAPG-ROM). While larger than the G-ROM, the eAPG-ROM maintains computational efficiency superior to the general APG-ROM formulation. Additionally, we introduce a novel, straightforward, yet efficient approach aimed at augmenting the number of degrees of freedom in the evaluation of the memory length, a critical factor influencing the stability and accuracy of the APG-ROM. This will involve a data-driven methodology to determine this augmented memory length.

We will present numerical results for the flow around a circular cylinder. The simulation will be conducted in 3D while considering turbulent flow. We will derive error measures for a comparative analysis of the numerical results. Furthermore, an evaluation of numerical computational costs will be conducted, considering metrics such as floating point operations (flops) and simulation time. This comprehensive analysis aims to provide an understanding of the trade-offs between computational efficiency and accuracy achieved by the eAPG-ROM.

The structure of the paper is as follows. 
Section~\ref{sec:G-ROM} outlines the standard G-ROM framework applied to fluid flows governed by the incompressible Navier-Stokes equations.
Section~\ref{sec:APG_ROM_general} describes the general APG-ROM, followed by the introduction of the eAPG-ROM framework in Section~\ref{sec:APG_ROM_NSE}. Additionally, we introduce the concept of an augmented memory length along with an optimization algorithm.
Section~\ref{sec:Error_Eval} introduces error measures used to evaluate the quality of the ROMs.
Section~\ref{sec:Implementation_and_computationalCost} gives details on the implementation and provides a comparison of computational costs.
Section~\ref{sec:numerical_example} presents the numerical experiment, showcasing the application and performance of the eAPG-ROM.
Section~\ref{sec:conclusion_and_outlook} summarizes the results with a brief conclusion and outlook.\\

\newpage
% \par\noindent\rule{\textwidth}{0.4pt}
\vspace*{-7mm}
\nomenclature[$a1$]{$\bm{a} \in\mathbb{R}^N$}{generalized coordinates}
\nomenclature[$a2$]{$\tilde{\bm{a}} \in\mathbb{R}^r$}{modal coefficients for the coarse-scale}
\nomenclature[$a3$]{$\tilde{\bm{a}}^{\text{POD}} \in\mathbb{R}^r$}{POD modal coefficients for the coarse-scale}
\nomenclature[$a4$]{$\tilde{\bm{a}}^{\text{G}} \in\mathbb{R}^r$}{G-ROM modal coefficients for the coarse-scale}
\nomenclature[$a5$]{$\tilde{\bm{a}}^{\text{APG}} \in\mathbb{R}^r$}{general APG-ROM modal coefficients for the coarse-scale}
\nomenclature[$a6$]{$\tilde{\bm{a}}^{\text{eAPG}} \in\mathbb{R}^r$}{eAPG-ROM modal coefficients for the coarse-scale}
\nomenclature[$a7$]{$\tilde{\bm{a}}^{\text{eAPG},\tau} \in\mathbb{R}^r$}{eAPG-ROM modal coefficients for the coarse-scale with arbitrarily chosen scalar memory length}
\nomenclature[$a8$]{$\tilde{\bm{a}}^{\text{eAPG},\tau,\text{opt}} \in\mathbb{R}^r$}{eAPG-ROM modal coefficients for the coarse-scale with optimized scalar memory length}
\nomenclature[$a9$]{$\tilde{\bm{a}}^{\text{eAPG},\bm{T},\text{opt}} \in\mathbb{R}^r$}{eAPG-ROM modal coefficients for the coarse-scale with optimized matrix memory length}
\nomenclature[$d1$]{$d \in\mathbb{R}$}{number of dimensions of the spatial domain}
\nomenclature[$d2$]{$dx$, $dy$, $dz$}{cell dimensions in x, y and z directions}
\nomenclature[$p$]{$\bm{p} \in\mathbb{R}^{N_{\text{grid}}}$}{pressure}
\nomenclature[$r$]{$r \in\mathbb{R}$}{number of reduced/coarse-scale base vectors}
\nomenclature[$t$]{$\Delta t \in\mathbb{R}$}{time step for numerical integration}
\nomenclature[$t$]{$t_m \in\mathbb{R}$}{discrete time}
\nomenclature[$u$]{$\bm{u} \in\mathbb{R}^N$}{velocity}
\nomenclature[$u$]{$\bm{u}^* \in\mathbb{R}^N$}{time-variant velocity}
\nomenclature[$u$]{$\bm{u}' \in\mathbb{R}N$}{time-averaged velocity}
\nomenclature[$u$]{$\tilde{\bm{u}} \in\mathbb{R}^N$}{velocity recovered with the coarse-scale base vectors}
\nomenclature[$w$]{$w \in\mathbb{R}$}{memory length weight}
\nomenclature[$x$]{$x_n \in\mathbb{R}$}{discrete spatial location}

\nomenclature[$C1$]{$\bm{C}^{\text{G},N} \in\mathbb{R}^{N}$}{constant coefficients of the G-ROM in $\mathbb{R}^N$}
\nomenclature[$C2$]{$\bm{C}^{\text{G}} \in\mathbb{R}^{r}$}{constant coefficients of the G-ROM}
\nomenclature[$C3$]{$\bm{C}^{\bm{\Pi}} \in\mathbb{R}^{N}$}{orthonormally projected constant coefficients of the G-ROM}
\nomenclature[$C4$]{$\bm{C}^{\text{APG},N} \in\mathbb{R}^{N}$}{constant coefficients of the efficient APG-ROM in $\mathbb{R}^N$}
\nomenclature[$C5$]{$\bm{C}^{\text{APG}} \in\mathbb{R}^{r}$}{constant coefficients of the efficient APG-ROM}
\nomenclature[$E$]{$\mathcal{E}_{\text{TRU}} \in\mathbb{R}$}{truncation error}
\nomenclature[$E$]{$\mathcal{E}_{\text{ROM}} \in\mathbb{R}$}{error introduced by the reduced order model}
\nomenclature[$E$]{$\mathcal{E}_{\text{Total}} \in\mathbb{R}$}{total error}
\nomenclature[$E$]{$\mathcal{E}_{\text{REC}} \in\mathbb{R}$}{reconstruction error}
\nomenclature[$F$]{$flops_{\text{G,Offline}} \in\mathbb{R}$}{flop count for the G-ROM Offline-Phase}
\nomenclature[$F$]{$flops_{\text{G,Online}} \in\mathbb{R}$}{flop count for the G-ROM Online-Phase}
\nomenclature[$F$]{$flops_{\text{eAPG,Offline}} \in\mathbb{R}$}{flop count for the efficient APG-ROM Offline-Phase}
\nomenclature[$F$]{$flops_{\text{eAPG,Online}} \in\mathbb{R}$}{flop count for the efficient APG-ROM Online-Phase}
\nomenclature[$F$]{$flops_{\text{APG}} \in\mathbb{R}$}{flop count for the General APG-ROM}
\nomenclature[$J$]{$\bm{J}(\cdot) \in\mathbb{R}$}{jacobian}
\nomenclature[$K1$]{$\bm{K}^{\text{APG},N} \in\mathbb{R}^{N\times r^3}$}{cubic coefficients of the efficient APG-ROM in $\mathbb{R}^N$}
\nomenclature[$K2$]{$\bm{K}^{\text{APG}} \in\mathbb{R}^{r\times r^3}$}{cubic coefficients of the efficient APG-ROM}
\nomenclature[$L1$]{$\bm{L}^{\text{G},N} \in\mathbb{R}^{N\times r}$}{linear coefficients of the G-ROM in $\mathbb{R}^N$}
\nomenclature[$L2$]{$\bm{L}^{\text{G}} \in\mathbb{R}^{r\times r}$}{linear coefficients of the G-ROM}
\nomenclature[$L3$]{$\bm{L}^{\bm{\Pi}} \in\mathbb{R}^{N\times r}$}{orthonormally projected linear coefficients of the G-ROM}
\nomenclature[$L4$]{$\bm{L}^{\text{APG},N} \in\mathbb{R}^{N\times r}$}{linear coefficients of the efficient APG-ROM in $\mathbb{R}^N$}
\nomenclature[$L5$]{$\bm{L}^{\text{APG}} \in\mathbb{R}^{r\times r}$}{linear coefficients of the efficient APG-ROM}
\nomenclature[$L6$]{$\bm{\mathcal{L}}$}{Liouville operator}
\nomenclature[$M1$]{$M \in\mathbb{R}$}{number of snapshots}
\nomenclature[$M2$]{$\bm{\mathcal{M}}$}{memory term}
\nomenclature[$N1$]{$N$}{$=dN_{\text{grid}} \in\mathbb{R}$}
\nomenclature[$N2$]{$N_{\text{grid}} \in\mathbb{R}$}{number of spatial locations on the grid}
\nomenclature[$Q1$]{$\bm{Q}^{\text{G},N} \in\mathbb{R}^{N\times r^2}$}{quadratic coefficients of the G-ROM in $\mathbb{R}^N$}
\nomenclature[$Q2$]{$\bm{Q}^{\text{G}} \in\mathbb{R}^{r\times r^2}$}{quadratic coefficients of the G-ROM}
\nomenclature[$Q3$]{$\bm{Q}^{\bm{\Pi}} \in\mathbb{R}^{N\times r^2}$}{orthonormally projected quadratic coefficients of the G-ROM}
\nomenclature[$Q4$]{$\bm{Q}^{\text{APG},N} \in\mathbb{R}^{N\times r^2}$}{quadratic coefficients of the efficient APG-ROM in $\mathbb{R}^N$}
\nomenclature[$Q5$]{$\bm{Q}^{\text{APG}} \in\mathbb{R}^{r\times r^2}$}{quadratic coefficients of the efficient APG-ROM}
\nomenclature[$R1$]{$\bm{R}(\cdot) \in\mathbb{R}^{N}$}{right-hand side operator}
\nomenclature[$R2$]{$R_e$}{reynolds number}
\nomenclature[$T$]{$T_{p} \in\mathbb{R}$}{time of the simulation for one flow period}
\nomenclature[$U1$]{$\bm{U}^* \in\mathbb{R}^{N\times M}$}{snapshot matrix of the time-variant velocity contributions}
\nomenclature[$V$]{$\bm{V} \in\mathbb{R}^{M\times M}$}{right-hand side eigenvector matrix of the POD}
\nomenclature[$W$]{$\bm{W} \in\mathbb{R}^{r\times r}$}{matrix memory length weighting matrix}

\nomenclature[A11]{$\Delta$, $\bm{\Delta}$}{spatial second differential operator}
\nomenclature[A121]{$\tilde{\bm{\Pi}}$}{resolved coarse-scale projection operator}
\nomenclature[A122]{$\bar{\bm{\Pi}}$}{unresolved fine-scale projection operator}
\nomenclature[A13]{$\bm{\Sigma} \in\mathbb{R}^{M\times M}$, $\sigma \in\mathbb{R}$}{singular values}
\nomenclature[A14]{$\bm{T} \in\mathbb{R}^{r\times r}$}{matrix memory length}
\nomenclature[A15]{$\bm{T}^{\text{opt}} \in\mathbb{R}^{r\times r}$}{optimized matrix memory length}
\nomenclature[A16]{$\bm{\Phi} \in\mathbb{R}^{N\times M}$, $\bm{\phi} \in\mathbb{R}^{N}$, $\phi \in\mathbb{R}^{d}$}{trial base vectors, POD modes}
\nomenclature[A17]{$\tilde{\bm{\Phi}} \in\mathbb{R}^{N\times r}$, $\tilde{\bm{\phi}} \in\mathbb{R}^{N}$, $\tilde{\phi} \in\mathbb{R}^{d}$}{coarse-scale trial base vectors, coarse-scale POD modes}
\nomenclature[A18]{$\bar{\bm{\Phi}} \in\mathbb{R}^{N\times r}$, $\bar{\bm{\phi}} \in\mathbb{R}^{N}$, $\bar{\phi} \in\mathbb{R}^{d}$}{fine-scale trial base vectors, fine-scale POD modes}
\nomenclature[A19]{$\bm{\Psi} \in\mathbb{R}^{N\times M}$, $\bm{\psi} \in\mathbb{R}^{N}$, $\psi \in\mathbb{R}^{d}$}{test base vectors}
\nomenclature[A21]{$\bm{\Psi}^{\text{APG}} \in\mathbb{R}^{N\times r}$}{adjoint Petrov-Galerkin test base vectors}

\nomenclature[A31]{$\delta_{ik}$}{Kronecker delta}
\nomenclature[A32]{$\nabla$, $\bm{\nabla}$}{spatial first differential operator}
\nomenclature[A34]{$\nu \in\mathbb{R}$}{kinematic viscosity}
\nomenclature[A35]{$\rho(\cdot) \in\mathbb{R}$}{spectral radius}
\nomenclature[A36]{$\tau \in\mathbb{R}$}{memory length}
\nomenclature[A37]{$\tau^{\text{opt}} \in\mathbb{R}$}{optimized memory length}
\nomenclature[A38]{$\bm{\varphi}(\cdot)$}{resolved, coarse modal coefficient operator}
\nomenclature[A39]{$\omega_1\in\mathbb{R}$, $\omega_2\in\mathbb{R}$}{Flop count for spatial differential operations}

\nomenclature[B]{FOM}{full order model}
\nomenclature[B]{POD}{proper orthogonal decomposition}
\nomenclature[B]{GP}{Galerkin projection}
\nomenclature[B]{SVD}{singular value decomposition}
\nomenclature[B]{ROM}{reduced order model}
\nomenclature[B]{G-ROM}{Galerkin reduced order model}
\nomenclature[B]{APG-ROM}{adjoint Petrov-Galerkin reduced order model}
\nomenclature[B]{eAPG-ROM}{efficient adjoint Petrov-Galerkin reduced order model}
\nomenclature[B]{Flops}{floating point operations}
\nomenclature[B]{SST}{shear-stress transport}
\nomenclature[B]{URANS}{unsteady Reynolds-Averaged Navier-Stokes}

% \begin{figure*}
%     \ContinuedFloat
%     \begin{mdframed}
%         \printnomenclature[2.7cm]
%     \end{mdframed}
% \end{figure*}

% \begin{tcolorbox}[%
%     enhanced, 
%     breakable,
%     skin first=enhanced,
%     skin middle=enhanced,
%     skin last=enhanced,
%     ]{}
            \printnomenclature[2.7cm]
            % \lipsum[1-15]

% \end{tcolorbox}

\vspace{-5mm}
\par\noindent\rule{\textwidth}{0.4pt}

\section*{Notation}\label{sec:notation}
Matrices are denoted by bold uppercase letters, vectors by bold lowercase letters, and scalars by lowercase letters. Functions represented by bold letters followed by parentheses indicate matrix or vector functions.
Bold spatial differential operators $\bm{\Delta}$ and $\bm{\nabla}$ apply the $\Delta$ and $\nabla$ operators to every $d$-dimensional Euclidean space vector. Consider, for example, a column vector	$\bm{g} = \left(\begin{matrix} \bm{g}(x_1)^T & \cdots & \bm{g}(x_{N_\text{grid}})^T \end{matrix}\right)^T\in\mathbb{R}^N$, with $\bm{g}(x_n)\in\mathbb{R}^d$, $n = 1,\hdots, N_\text{grid}$, where $N_{\text{grid}}$ is the total number of spatial points on a discrete spatial grid, $d=3$ throughout the paper and $N = dN_{\text{grid}}$. Then $\bm{\nabla} \bm{g} = \left(\begin{matrix} \nabla \bm{g}(x_1)^T & \cdots & \nabla \bm{g}(x_{N_\text{grid}})^T\end{matrix}\right)^T$ and $\bm{\Delta} \bm{g}= \left(\begin{matrix}\Delta \bm{g}(x_1)^T & \cdots & \Delta \bm{g}(x_{N_\text{grid}})^T \end{matrix}\right)^T$.

\section{POD-Galerkin Reduced Order Model for fluid flows}\label{sec:G-ROM}
Our objective is to implement a projection-based reduced order model for fluid flows governed by the incompressible Navier-Stokes equations,
\begin{subequations}\label{eq:NSE}
	\begin{align}
		\frac{\partial}{\partial t} \bm{u}(t) &= -(\bm{u}(t)\cdot \bm{\nabla})\bm{u}(t) +\nu\bm{\Delta} \bm{u}(t) -\bm{\nabla} \bm{p}(t),\label{eq:NSE1}
		\\
		\bm{\nabla} \cdot \bm{u}(t) &= \bm{0},
		\label{eq:NSE2}
		\\
		\bm{u}(0) &= \bm{u}_0,\quad
		\bm{p}(0) = \bm{p}_0,\label{eq:NSE3}
	\end{align}
\end{subequations}
where $\bm{u}(t)=\left[\begin{matrix}\bm{u}(x_1,t)^T & \hdots & \bm{u}(x_{N_{\text{grid}}},t)^T\end{matrix}\right]^T \in \mathbb{R}^N$ and $\bm{p}(t) = \left[\begin{matrix}p(x_1,t) & \hdots & p(x_{N_{\text{grid}}},t)\end{matrix}\right]^T \in \mathbb{R}^{N_{\text{grid}}}$ denote the fluid velocities and pressures of the $N$-dimensional dynamical system with $\bm{u}(x_n,t) \in \mathbb{R}^{d}$ and $\bm{p}(x_n,t) \in \mathbb{R}$, $n=1,\hdots,N_{\text{grid}}$, respectively, with initial system states $\bm{u}_0 \in \mathbb{R}^N$ and $\bm{p}_0 \in \mathbb{R}^{N_{\text{grid}}}$. We abbreviate \eqref{eq:NSE1} by
\begin{equation}\label{eq:FOM}
	\begin{aligned}
		\dot{\bm{u}}(t) = \bm{R}(\bm{u}(t)),
	\end{aligned}
\end{equation}
and neglect the pressure at this stage (see, e.g., \cite{John2010}).

\subsection{Weak form of the FOM}\label{sec:weak_Form}
Let $\bm{\Phi} = \left[\begin{matrix} \bm{\phi}_1 & \hdots & \bm{\phi}_N \end{matrix}\right]\in \mathbb{R}^{N\times N}$ denote a matrix that collects $N$ orthonormal basis vectors for $\mathbb{R}^{N}$, where $\bm{\phi}_k=\left[\begin{matrix}\bm{\phi}_k(x_1)^T & \hdots & \bm{\phi}_k(x_{N_{\text{grid}}})^T\end{matrix}\right]^T \in \mathbb{R}^N, k=1,\hdots, N$ and $\bm{\phi}_k(x_n) \in \mathbb{R}^d$, $n=1,\hdots, N_{\text{grid}}$. Whenever such a collection of basis vectors is used to represent the velocity in the form 
\begin{align}\label{eq:FOM_exact_approx}
	\bm{u}(t) = \sum_{i=1}^N \bm{\phi}_i a_i(t) = \bm{\Phi} \bm{a}(t),
\end{align}
where $\bm{a}(t) = \left[\begin{matrix}	a_1(t) & \hdots & a_N(t)\end{matrix}\right]^T\in \mathbb{R}^N$, we call $\bm{\Phi}$ and the basis vectors it collects a trial basis. In \eqref{eq:FOM_exact_approx}, $\bm{a}(t)$ are the $N$-dimensional state variables, which we call generalized coordinates. Similarly, we define a test basis $\bm{\Psi} = \left[\begin{matrix}	\bm{\psi}_1 & \hdots & \bm{\psi}_N\end{matrix}\right]\in \mathbb{R}^{N\times N}$, where $\bm{\psi}_k=\left[\begin{matrix}\bm{\psi}_k(x_1)^T & \hdots & \bm{\psi}_k(x_{N_{\text{grid}}})^T\end{matrix}\right]^T \in \mathbb{R}^N, k=1,\hdots, N$ and $\bm{\psi}_k(x_n) \in \mathbb{R}^d$, $n=1,\hdots, N_{\text{grid}}$. Substituting~\eqref{eq:FOM_exact_approx} into~\eqref{eq:FOM} and left multiplying by $\bm{\Psi}^T$ yields the projected FOM expressed with generalized coordinates, $\bm{\Psi}^T\bm{\Phi}\dot{\bm{a}}(t) = \bm{\Psi}^T\bm{R}(\bm{\Phi a}(t))$, or equivalently 
\begin{align}\label{eq:FOM_weak_Form}
	\dot{\bm{a}}(t) &= (\bm{\Psi}^T\bm{\Phi})^{-1}\bm{\Psi}^T\bm{R}(\bm{\Phi a}(t)).
\end{align}
The corresponding treatment of the initial condition~\eqref{eq:NSE3} yields	$\bm{a}(0) = \bm{a}_0 = (\bm{\Psi}^T\bm{\Phi})^{-1}\bm{\Psi}^T \bm{u}_0$.
Equation~\eqref{eq:FOM_weak_Form} is often called the weak form of the FOM because it involves a projection onto the test basis, thereby relaxing the smoothness requirements on the solution (see, e.g., \cite{Hughes2000}).

\subsection{Proper Orthogonal Decomposition of Snapshot-Data}\label{sec:POD}
It remains to explain how to construct the test and trial basis. We will focus on constructing the trial basis $\bm{\Phi}$ first. A common method is to use the POD of snapshot-data~\cite{Sirovich1987a}. 
We solve the FOM~\eqref{eq:NSE} for this purpose and obtain a sampled solution $\bm{u}(t_m) \in \mathbb{R}^N$ for time steps $t_m$, $m = 1,\hdots,M$, where $M$ denotes the total number of steps or snapshots. 
After separating the time average $\bm{u}' = \frac{1}{M}\sum_{m=1}^{M}\bm{u}(t_m)$ according to $\bm{u}(t_m) = \bm{u}' + \bm{u}^*(t_m)$, we construct the snapshot matrix $\bm{U}^* = \left[\begin{matrix} \bm{u}^*(t_1) & \hdots & \bm{u}^*(t_M) \end{matrix}\right]\in \mathbb{R}^{N\times M}$, where $M\ll N$ and $\text{rank}(\bm{U}^*) = M$. We compute the trial basis vectors with a singular value decomposition (SVD), $\bm{U}^* = \bm{\Phi} \Sigma \bm{V}^T$, and refer to the columns of $\bm{\Phi}$ as the POD modes. Since $M\ll N$, the singular values $\sigma_k = 0$ for $k=M+1,\hdots,N$ in $\bm{\Sigma} = \text{diag}(\sigma_1, \hdots, \sigma_M)$, so $\bm{\phi}_k$ for $k=M+1,\hdots,N$ can be omitted. Equivalently, a thin SVD (see, e.g.,~\cite[Chapter~6]{Zhaojun2000}) can be carried out to obtain $\bm{\Phi} = \left[\begin{matrix} \bm{\phi}_1 & \cdots & \bm{\phi}_M	\end{matrix}\right]\in \mathbb{R}^{N\times M}$, the thin singular value matrix $\bm{\Sigma} \in\mathbb{R}^{M\times M}$ and $\bm{V}\in \mathbb{R}^{M\times M}$. Equation~\eqref{eq:FOM_exact_approx} can be reformulated to read
\begin{align*}
	\bm{u}(t) = \bm{u}' + \bm{u}^*(t) = \bm{u}' + \sum_{i=1}^M \bm{\phi}_i a_i(t) = \bm{u}' + \bm{\Phi} \bm{a}(t).
\end{align*}
We refer to $\bm{a}(t) = \left[\begin{matrix}a_1(t) & \hdots & a_M(t) \end{matrix}\right]^T\in \mathbb{R}^{M}$ as modal coefficients.

\subsection{Basis truncation}\label{sec:Base_truncation}
In order to construct a ROM we seek a low-dimensional representation of the FOM, where the number of reduced dimensions satisfies $r\ll N$. We reformulate~\eqref{eq:FOM_weak_Form} with a multiscale ansatz that distinguishes between resolved and unresolved contributions. The theoretical justification for this approach will be elaborated in Section~\ref{sec:APG_ROM_general}. We split the space spanned by the trial basis vectors $\bm{\phi}_k$, $k = 1,\hdots,M$ in a resolved coarse-scale subspace spanned by the first $r$ trial basis vectors $\tilde{\bm{\phi}}_k$, $k=1,\hdots,r$ and an unresolved fine-scale subspace spanned by the remaining $\bar{\bm{\phi}}_k$, $k=1,\hdots,M-r$, where $\tilde{\bm{\Phi}} = \left[\begin{matrix} \bm{\phi}_1 & \cdots & \bm{\phi}_r \end{matrix}\right]$ and $\bar{\bm{\Phi}} = \left[\begin{matrix} \bm{\phi}_{r+1} & \cdots & \bm{\phi}_M \end{matrix}\right]$.
We refer to $\tilde{\bm{\phi}}_k$, $k=1,\hdots,r$ as the resolved coarse-scale trial basis vectors and to $\bar{\bm{\phi}}_k$, $k=1,\hdots,M-r$ as the unresolved fine-scale trial basis vectors. 
We want to derive a model capable of capturing the dynamics of the FOM while operating with a reduced dimensionality, focusing on the resolved coarse scales. This involves the truncation of the basis vectors. The effect of this truncation can be managed by selecting an appropriate value for $r$, allowing control over the truncation error given by
\begin{align}\label{sec:TruncationError}
	\mathcal{E}_{\text{Tru}}(r) = 1 - \dfrac{\sum\limits_{k=1}^{r}\sigma_k^2}{\sum\limits_{k=1}^{M}\sigma_k^2},
\end{align}
and ensuring it remains sufficiently small.
The singular values $\sigma_k$, $k=1,\hdots,M$ are ordered, i.e., $\sigma_1 \geq \sigma_2 \geq \hdots \geq \sigma_M$.
Similarly, we partition the resolved and unresolved modal coefficients, where $\tilde{\bm{a}}(t) = \left[\begin{matrix} a_1(t) & \hdots & a_r(t) \end{matrix}\right]^T$ and $\bar{\bm{a}}(t) = \left[\begin{matrix} a_{r+1}(t) & \hdots & a_M(t) \end{matrix}\right]^T$.

\subsection{Galerkin Projection}\label{sec:Galerkin_Projection}
We decompose the time-variant velocity $\bm{u}^*(t) = \tilde{\bm{u}}^*(t) + \bar{\bm{u}}^*(t)$ into two components: $\tilde{\bm{u}}^*(t)$ representing the resolved contributions and $\bar{\bm{u}}^*(t)$ representing the unresolved contributions, where
\begin{align}
	\tilde{\bm{u}}^*(t) = \tilde{\bm{\Phi}} \tilde{\bm{a}}(t),\label{eq:identity_resolved_velocity}\\
	\bar{\bm{u}}^*(t) = \bar{\bm{\Phi}} \bar{\bm{a}}(t).\nonumber
\end{align}
We introduce two projection operators: The resolved coarse-scale projection operator $\tilde{\bm{\Pi}} = \tilde{\bm{\Phi}}\tilde{\bm{\Phi}}^T$ and the unresolved fine-scale projection operator $\bar{\bm{\Pi}} = \bm{I} - \tilde{\bm{\Pi}}$. These operators map from the full space to the resolved space and the unresolved space, respectively. The projection operators satisfy $\tilde{\bm{\Pi}}^2 = \tilde{\bm{\Pi}}$, $\bar{\bm{\Pi}}^2 = \bar{\bm{\Pi}}$, $\tilde{\bm{\Pi}} + \bar{\bm{\Pi}} = \bm{I}$, and $\tilde{\bm{\Pi}} \bar{\bm{\Pi}} = 0$.
We apply the coarse-scale projection operator $\tilde{\bm{\Pi}}$ to project the FOM~\eqref{eq:FOM} onto the resolved coarse scales.
For standard Galerkin ROMs the influence of the unresolved fine scales is usually neglected (see \ref{ap:Coarse-scale_Projection} for a brief explanation) and \eqref{eq:FOM} becomes
\begin{align}\label{eq:FOM_Projected_divided}
	\dot{\tilde{\bm{u}}}^*(t)  := \tilde{\bm{\Pi}}\bm{R}(\bm{u}' + \tilde{\bm{u}}^*(t)).
\end{align}
Substituting~\eqref{eq:identity_resolved_velocity} into~\eqref{eq:FOM_Projected_divided} and projecting the resulting equation onto the test basis yields $\bm{\Psi}^T \tilde{\bm{\Phi}} \dot{\tilde{\bm{a}}}(t)  = \bm{\Psi}^T \tilde{\bm{\Pi}}\bm{R}(\bm{u}' + \tilde{\bm{\Phi}} \tilde{\bm{a}}(t))$. For the Galerkin projection we choose $\bm{\Psi} = \tilde{\bm{\Phi}}$. This results in a set of $r$ ODEs
\begin{equation}\label{eq:G_ROM_with_rhsOperator}
	\begin{aligned}
		\dot{\tilde{\bm{a}}}(t) &= \tilde{\bm{\Phi}}^T\tilde{\bm{\Phi}}\tilde{\bm{\Phi}}^T \bm{R}(\bm{u}' + \tilde{\bm{\Phi}} \tilde{\bm{a}}(t))
		= \tilde{\bm{\Phi}}^T \bm{R}(\bm{u}' + \tilde{\bm{\Phi}} \tilde{\bm{a}}(t)).
	\end{aligned}
\end{equation}
Once the modal coefficients $\tilde{\bm{a}}(t)$ are known from solving~\eqref{eq:G_ROM_with_rhsOperator} the full order state can be reconstructed with 
\begin{align}\label{eq:FOM_exact_approx_in_R}
	\tilde{\bm{u}}(t) = \bm{u}' + \tilde{\bm{u}}^*(t) = \bm{u}' + \sum_{i=1}^M \tilde{\bm{\phi}}_i \tilde{a}_i(t) = \bm{u}' + \tilde{\bm{\Phi}} \tilde{\bm{a}}(t).
\end{align}

\subsection{Galerkin Reduced Order Model for fluid flows}\label{sec:G_ROM_NSE}
To derive a ROM for~\eqref{eq:NSE}, we start by examining $\bm{R}(\tilde{\bm{u}}(t)) = \bm{R}(\bm{u}' + \tilde{\bm{\Phi}} \tilde{\bm{a}}(t))$. It is common practice to neglect the pressure term (see, e.g.,~\cite{John2010}) and~\eqref{eq:NSE} becomes $\bm{R}(\bm{u}' + \tilde{\bm{\Phi}} \tilde{\bm{a}}(t)) = -((\bm{u}' + \tilde{\bm{\Phi}} \tilde{\bm{a}}(t))\cdot \bm{\nabla})(\bm{u}' + \tilde{\bm{\Phi}} \tilde{\bm{a}}(t)) + \nu\bm{\Delta} (\bm{u}' + \tilde{\bm{\Phi}} \tilde{\bm{a}}(t))$.
Organizing this equation by powers of the modal coefficients provides
\begin{align*}
	\bm{R}(\bm{u}' + \tilde{\bm{\Phi}} \tilde{\bm{a}}(t)) =
		-(\tilde{\bm{\Phi}} \tilde{\bm{a}}(t)\cdot \bm{\nabla})\tilde{\bm{\Phi}} \tilde{\bm{a}}(t)
		-(\tilde{\bm{\Phi}} \tilde{\bm{a}}(t)\cdot \bm{\nabla})\bm{u}'
		-(\bm{u}'\cdot \bm{\nabla})\tilde{\bm{\Phi}} \tilde{\bm{a}}(t)
		+ \nu\bm{\Delta} \tilde{\bm{\Phi}} \tilde{\bm{a}}(t)
		-(\bm{u}'\cdot \bm{\nabla})\bm{u}'
		+ \nu\bm{\Delta} \bm{u}'.
\end{align*}
Since the modal coefficients possess no spatial dependence, we can rearrange this equation to expose its quadratic nature
\begin{equation}
	\begin{aligned}\label{eq:G_ROM_NS_Arranged_notProjected}
		\bm{R}(\bm{u}' + \tilde{\bm{\Phi}} \tilde{\bm{a}}(t)) = 
			\bm{Q}^{\text{G},N} \big[\tilde{\bm{a}}(t)\otimes \tilde{\bm{a}}(t)\big] 
			+ \bm{L}^{\text{G},N} \tilde{\bm{a}}(t) 
			+ \bm{C}^{\text{G},N},
	\end{aligned}
\end{equation}
with
\begin{equation}
	\begin{aligned}\label{eq:G_ROM_NS_Arranged_notProjected_coefficients}
		\bm{Q}^{\text{G},N} &= -(\tilde{\bm{\Phi}} \cdot \bm{\nabla})\tilde{\bm{\Phi}} \in\mathbb{R}^{N\times r^2},\\
		\bm{L}^{\text{G},N} &= -(\tilde{\bm{\Phi}} \cdot \bm{\nabla})\bm{u}'
		-(\bm{u}'\cdot \bm{\nabla})\tilde{\bm{\Phi}} 
		+ \nu\bm{\Delta} \tilde{\bm{\Phi}} \in\mathbb{R}^{N\times r},\\
		\bm{C}^{\text{G},N} &= -(\bm{u}'\cdot \bm{\nabla})\bm{u}'
		+ \nu\bm{\Delta} \bm{u}' \in\mathbb{R}^{N},
	\end{aligned}
\end{equation}
where $\otimes$ denotes the Kronecker product and $\tilde{\bm{a}}\otimes \tilde{\bm{a}} = \left(\begin{matrix} \tilde{a}_1\tilde{a}_1 & \cdots & \tilde{a}_1\tilde{a}_r & \cdots & \tilde{a}_r\tilde{a}_r \end{matrix}\right)^T \in \mathbb{R}^{r^2}$.
Upon rearranging the space-dependent components as detailed in~\ref{ap:Matrices_G_ROM} and projecting~\eqref{eq:G_ROM_NS_Arranged_notProjected} and~\eqref{eq:G_ROM_NS_Arranged_notProjected_coefficients} onto the coarse-scale trial basis $\tilde{\bm{\Phi}}$, we arrive at the formulation for the proper orthogonal decomposition Galerkin projection ROM (G-ROM) for fluid flows governed by the incompressible Navier-Stokes equations. The G-ROM is expressed by the following set of ordinary differential equations
\begin{subequations}\label{eq:G_ROM}
	\begin{align}
		\begin{split}\label{eq:G_ROM_eq}
			\dot{\tilde{\bm{a}}}^{\text{G}}(t) 
			&=
				\tilde{\bm{\Phi}}^T \left(\bm{Q}^{\text{G},N}\big[\tilde{\bm{a}}^\text{G}(t)\otimes \tilde{\bm{a}}^\text{G}(t)\big] 
				+ \bm{L}^{\text{G},N} \tilde{\bm{a}}^\text{G}(t) 
				+ \bm{C}^{\text{G},N}\right)\\
			&=
				\bm{Q}^{\text{G}} \big[\tilde{\bm{a}}^\text{G}(t)\otimes \tilde{\bm{a}}^G(t)\big] 
				+ \bm{L}^{\text{G}} \tilde{\bm{a}}^\text{G}(t) 
				+ \bm{C}^{\text{G}},
		\end{split}\\
		\begin{split}\label{eq:G_ROM_initialCondition}
			\tilde{\bm{a}}^\text{G}(0) &= \tilde{\bm{a}}_0 = \tilde{\bm{\Phi}}^T \bm{u}^*_0,
		\end{split}
	\end{align}
\end{subequations}
where
\begin{equation}\label{eq:G_ROM_coeffs}
	\begin{aligned}
		\bm{Q}^{\text{G}} &= \tilde{\bm{\Phi}}^T \bm{Q}^{\text{G},N}\in\mathbb{R}^{r\times r^2},\\
		\bm{L}^{\text{G}} &= \tilde{\bm{\Phi}}^T \bm{L}^{\text{G},N}\in\mathbb{R}^{r\times r},\\
		\bm{C}^{\text{G}} &= \tilde{\bm{\Phi}}^T \bm{C}^{\text{G},N}\in\mathbb{R}^{r}.
	\end{aligned}
\end{equation}
Note that in the ROM formulation, the continuity equation~\eqref{eq:NSE2} is omitted since the zero divergence of the velocity is inherently guaranteed by the CFD simulation data, simplifying the model without sacrificing accuracy (see, e.g.,~\cite{John2010}). The initial condition is determined by projecting the initial condition of the FOM onto the coarse-scale trial basis. The solution of the G-ROM \eqref{eq:G_ROM} is denoted $\tilde{\bm{a}}^\text{G}(t)$. Subsequently, the velocity field is reconstructed with~\eqref{eq:FOM_exact_approx_in_R}.

%To streamline the mathematical operations and enhance the clarity of the method, we rearrange and vectorize the operations involved. This not only results in a concise and elegant expression for the G-ROM but also facilitates a straightforward implementation. Additionally, as will be demonstrated later, this vectorized representation proves instrumental in simplifying the derivation of the efficient Adjoint Petrov-Galerkin Reduced Order Model for the incompressible Navier-Stokes equations.

The G-ROM can be distinctly divided into two phases: the computationally expensive offline phase and the comparatively efficient online phase. In the offline phase, numerous mathematical operations are performed in the spatial domain $\mathbb{R}^N$, where $N$ may become very large. This phase involves the derivation of the ROM~\eqref{eq:G_ROM_eq} and the computation of its associated coefficients~\eqref{eq:G_ROM_coeffs}. The offline phase is resource-intensive and typically requires substantial computational power. However, once the G-ROM and its coefficients have been calculated in the offline phase, the online phase becomes significantly less computationally demanding due to the reduced dimensionality $r\ll N$. This clear distinction between the offline and the online phase is a characteristic feature of projection-based ROMs for fluid flows, allowing for substantial computational savings during the runtime of simulations. The ease of use and implementation of the standard G-ROM make it an attractive choice for modeling fluid flow governed by the Navier-Stokes equations.

\section{Adjoint Petrov-Galerkin Reduced Order Model}\label{sec:APG_ROM_general}
\begin{figure}[b!]
	\centering
	\def\svgwidth{\columnwidth}
	\includegraphics{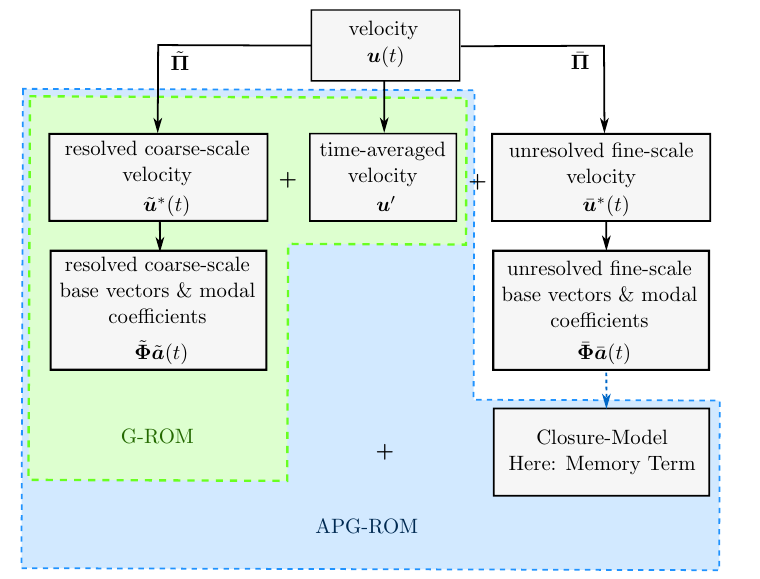}
	\captionsetup{width=\linewidth}
	\caption{Schematic representation of the resolved coarse-scale and unresolved fine-scale components in the ROM. The blue dashed line denotes the approximation of the influence of the unresolved fine scales on the resolved coarse-scale dynamics leading to the closure model, specifically the memory term discussed in Section~\ref{sec:APG_ROM_general}.} 
	\label{fig:Closure_Model}
\end{figure}

Due to nonlinearities in the FOM the fine scales in the right-hand side operator in~\eqref{eq:FOM_Projected_divided} cannot be neglected. Because the unresolved fine scales inevitably influence the resolved scales, \eqref{eq:FOM_Projected_divided} becomes	$\dot{\tilde{\bm{u}}}^*(t) - \tilde{\bm{\Pi}}\bm{R}(\bm{u}' + \tilde{\bm{u}}^*(t)) \neq 0$.
Consequently, a correction is needed to achieve an accurate representation of the FOM dynamics in the resolved space. 
This challenge is commonly referred to as the closure problem for ROMs (see, e.g.,~\cite{Ahmed2021, Couplet2003, Bergmann2009}), requiring the development of suitable closure models. A schematic overview for this problem is shown in Figure~\ref{fig:Closure_Model}. In summary, the standard G-ROM fails to capture the influence of the fine scales on the coarse-scale dynamics, leading to inaccuracies, particularly in highly turbulent flows (see, e.g.,~\cite{Akhtar2012, Baiges2015}). 

The Petrov-Galerkin method (see, e.g.,~\cite{Carlberg2011, Giere2015, Reineking2022, Parish2020}) emerged as a promising strategy to address the closure problem and to improve the overall performance of projection-based ROMs. This method involves carefully selecting trial and test bases, allowing for a more tailored and accurate representation of the underlying fluid dynamics. 
This section provides a brief outline of the APG-ROM method, as introduced in~\cite{Parish2020}. For a detailed explanation, we refer to~\cite{Parish2020} and~\cite{Chorin2009}. The APG-ROM is based on the Mori-Zwanzig Formalism~\cite{Zwanzig1973,Mori1965} tailored specifically for ROMs in the context of POD-GP and addresses the challenge of modeling the impact of unresolved fine scales on the resolved coarse scales without explicitly incorporating the fine scales, thus mitigating issues associated with ROM closure. 

\subsection{Liouville-Equation}\label{sec:Liouville-Equation}
The Mori-Zwanzig method for ROMs involves the transformation of the original FOM. Specifically, when dealing with incompressible Navier-Stokes equations, the FOM consists of nonlinear differential equations. Unfortunately, the Mori-Zwanzig approach cannot be directly employed for nonlinear problems. We address the nonlinearity by reinterpreting the nonlinear differential equations as equivalent linear partial differential equations. 
%This transformation leads to the generalized Mori-Zwanzig formalism.

The first step involves expressing the FOM solution $\bm{u}(t)$ in the weak form~\eqref{eq:FOM_weak_Form} as the sum of the time-averaged component $\bm{u}'$ and the product of the trial basis vectors $\bm{\Phi}$ and modal coefficients $\bm{a}(t)$, and then executing the Galerkin Projection, i.e., equating the trial and test bases $\bm{\Psi} = \bm{\Phi}$
\begin{align}
	\frac{\partial}{\partial t}\bm{a}(t) &= \bm{\Phi}^T \bm{R}(\bm{u}' + \bm{\Phi a}(t)).\label{eq:stateSpaceFOM_state}
\end{align}
We apply an operator denoted by $\bm{\varphi}: \mathbb{R}^M \rightarrow \mathbb{R}^r$ to the modal coefficients $\bm{a}(t)$. This operation yields the resolved coarse scale modal coefficients	$\tilde{\bm{a}}(t) = \bm{\varphi}(\bm{a}(t))$.
The solution of \eqref{eq:stateSpaceFOM_state} obviously depends on the initial condition $\bm{a}_0 = \bm{a}(0)$. Because we need derivatives of the solution w.r.t $\bm{a}_0$, we include the dependence on $\bm{a}_0$ explicitly, i.e.,
\begin{subequations}\label{eq:stateSpaceFOM_inital}
	\begin{align}
		\frac{\partial}{\partial t}\bm{a}(\bm{a}_0,t) &= \bm{\Phi}^T\bm{R}(\bm{u}' + \bm{\Phi a}(\bm{a}_0,t)),\label{eq:stateSpaceFOM_state_inital}\\
		\tilde{\bm{a}}(\tilde{\bm{a}}_0,t) &= \bm{\varphi}(\bm{a}(\bm{a}_0,t)),\label{eq:stateSpaceFOM_output_inital}
	\end{align}
\end{subequations}
and compute the time derivative of~\eqref{eq:stateSpaceFOM_output_inital}
\begin{equation}\label{eq:Lie_derivative}
	\begin{aligned}
		\frac{d}{dt}\tilde{\bm{a}}(\tilde{\bm{a}}_0,t) &= \frac{d}{dt}\bm{\varphi}(\bm{a}(\bm{a}_0,t))
			= \frac{\partial \bm{\varphi}(\bm{a}(\bm{a}_0,t))}{\partial \bm{a}(\bm{a}_0,t)} \frac{\partial}{\partial t}\bm{a}(\bm{a}_0,t)
			= \frac{\partial \bm{\varphi}(\bm{a}(\bm{a}_0,t))}{\partial \bm{a}(\bm{a}_0,t)} \bm{\Phi}^T\bm{R}(\bm{u}' + \bm{\Phi a}(t)).
	\end{aligned}
\end{equation}
It can be shown that (see~\cite[Chapter~6.3.1]{Chorin2009})
\begin{align}\label{eq:RHS_initalState}
	\bm{\Phi}^T\bm{R}(\bm{u}' + \bm{\Phi a}(t)) = \frac{\partial \bm{a}(\bm{a}_0,t)}{\partial \bm{a}_0} \bm{\Phi}^T\bm{R}(\bm{u}' + \bm{\Phi a}_0),
\end{align}
where \eqref{eq:RHS_initalState} is an exact representation without approximation.
Inserting~\eqref{eq:RHS_initalState} into~\eqref{eq:Lie_derivative} and using the chain rule yields
\begin{equation*}\label{eq:Lie_derivative2}
	\begin{aligned}
		\frac{d}{dt}\tilde{\bm{a}}(\tilde{\bm{a}}_0,t) &= \frac{\partial \bm{\varphi}(\bm{a}(\bm{a}_0,t))}{\partial \bm{a}(\bm{a}_0,t)} \frac{\partial \bm{a}(\bm{a}_0,t)}{\partial \bm{a}_0} \bm{\Phi}^T\bm{R}(\bm{u}' + \bm{\Phi a}_0)
		= \frac{d \bm{\varphi}(\bm{a}(\bm{a}_0,t))}{d \bm{a}_0} \bm{\Phi}^T\bm{R}(\bm{u}' + \bm{\Phi a}_0).\\
	\end{aligned}
\end{equation*}
This step leads to the Liouville equation for $\tilde{\bm{a}}(\tilde{\bm{a}}_0,t)$
\begin{align}\label{eq:Liouville-equation}
	\frac{\partial}{\partial t}\tilde{\bm{a}}(\tilde{\bm{a}}_0,t) = \bm{\mathcal{L}} \tilde{\bm{a}}(\tilde{\bm{a}}_0,t),
\end{align}
where the Liouville operator is defined as $\bm{\mathcal{L}}v = \frac{\partial}{\partial \bm{a}_0}v \bm{\Phi}^T\bm{R}(\bm{u}' + \bm{\Phi a}_0)$, for an arbitrary function $v$. Note that the total derivative with respect to time was replaced by a partial derivative in \eqref{eq:Liouville-equation}, because $\tilde{\bm{a}}_0$ is independent of time.
The Liouville equation~\eqref{eq:Liouville-equation} provides an exact linear representation of the original system \eqref{eq:stateSpaceFOM_state} \cite{Chorin2009}. This linearity enables us to derive the standard solution to~\eqref{eq:Liouville-equation}, given by $\tilde{\bm{a}}(\tilde{\bm{a}}_0,t) = e^{t\bm{\mathcal{L}}} \tilde{\bm{a}}_0$, where $e^{t\bm{\mathcal{L}}}$ is referred to as a propagator. This propagator evolves the solution from the initial condition $\tilde{\bm{a}}_0$ over time $t$. 
Inserting the propagator into~\eqref{eq:Liouville-equation}, we obtain
\begin{align}\label{eq:Liouville-equation_initial}
	\frac{\partial}{\partial t}e^{t\bm{\mathcal{L}}} \tilde{\bm{a}}_0 = e^{t\bm{\mathcal{L}}}\bm{\mathcal{L}}\tilde{\bm{a}}_0.
\end{align}

\subsection{Mori-Zwanzig Equation}\label{sec:Mori-Zwanzig-Equation}
Recalling the partitioning of the space into resolved and unresolved subspaces and the corresponding projection operators $\tilde{\bm{\Pi}}$ and $\bar{\bm{\Pi}}$, as discussed in Section~\ref{sec:Base_truncation}, we split the right-hand side of~\eqref{eq:Liouville-equation_initial} into
\begin{align}\label{eq:Liouville-equation_initial_split}
	\frac{\partial}{\partial t}e^{t\bm{\mathcal{L}}} \tilde{\bm{a}}_0 = e^{t\bm{\mathcal{L}}}\bm{I}\bm{\mathcal{L}}\tilde{\bm{a}}_0 = e^{t\bm{\mathcal{L}}}\tilde{\bm{\Pi}}\bm{\mathcal{L}}\tilde{\bm{a}}_0 + e^{t\bm{\mathcal{L}}}\bar{\bm{\Pi}}\bm{\mathcal{L}}\tilde{\bm{a}}_0.
\end{align}
Moreover, we can use the Duhamel Principle~\cite{Chorin2009} with the propagator, resulting in the decomposition
\begin{align}\label{eq:Liouville-equation_initial_split_Duhamel}
	e^{t(\tilde{\bm{\Pi}} + \bar{\bm{\Pi}})\bm{\mathcal{L}}} = e^{t\bar{\bm{\Pi}}\bm{\mathcal{L}}} + \int_0^t e^{(t-s)\bm{\mathcal{L}}} \tilde{\bm{\Pi}}\bm{\mathcal{L}} e^{s\bar{\bm{\Pi}}\bm{\mathcal{L}}} ds,
\end{align}
and apply the Duhamel formula~\eqref{eq:Liouville-equation_initial_split_Duhamel} to the second term on the right-hand side of~\eqref{eq:Liouville-equation_initial_split}, which yields the desired Mori-Zwanzig equation
\begin{align}\label{eq:Mori-Zwanzig-equation}
	\frac{\partial}{\partial t}e^{t\bm{\mathcal{L}}} \tilde{\bm{a}}_0 = e^{t\bm{\mathcal{L}}}\tilde{\bm{\Pi}}\bm{\mathcal{L}}\tilde{\bm{a}}_0 + e^{t\bar{\bm{\Pi}}\bm{\mathcal{L}}}\bar{\bm{\Pi}}\bm{\mathcal{L}}\tilde{\bm{a}}_0 + \int_0^t e^{(t-s)\bm{\mathcal{L}}} \tilde{\bm{\Pi}}\bm{\mathcal{L}} e^{s\bar{\bm{\Pi}}\bm{\mathcal{L}}} \bar{\bm{\Pi}}\bm{\mathcal{L}}\tilde{\bm{a}}_0ds.
\end{align}
The second term on the right-hand side is called the \textit{noise} term and becomes zero (see~\cite{Chorin2009,Parish2020}).
For simplicity, we now remove the explicit dependencies on the initial states as introduced in~\eqref{eq:stateSpaceFOM_inital} and use the identities $\frac{\partial}{\partial t}e^{t\bm{\mathcal{L}}} \tilde{\bm{a}}_0 = \frac{\partial}{\partial t}\tilde{\bm{a}}(t) = \frac{\partial}{\partial t}\tilde{\bm{\Phi}}^T\tilde{\bm{u}}^*(t)$ and $e^{t\bm{\mathcal{L}}}\tilde{\bm{\Pi}}\bm{\mathcal{L}}\tilde{\bm{a}}_0 = \tilde{\bm{\Phi}}^T \bm{R}(\tilde{\bm{u}}(t))$ for the terms in~\eqref{eq:Mori-Zwanzig-equation}~\cite{Parish2020}.
The Mori-Zwanzig equation~\eqref{eq:Mori-Zwanzig-equation} becomes
\begin{align}\label{eq:Mori-Zwanzig-equation2}
	\tilde{\bm{\Phi}}^T\left(\frac{\partial}{\partial t}\tilde{\bm{u}}^*(t) - \bm{R}(\tilde{\bm{u}}(t))\right) = \int_0^t e^{(t-s)\bm{\mathcal{L}}} \tilde{\bm{\Pi}}\bm{\mathcal{L}} e^{s\bar{\bm{\Pi}}\bm{\mathcal{L}}} \bar{\bm{\Pi}}\bm{\mathcal{L}}\tilde{\bm{a}}_0ds,
\end{align}
and retains an exact representation of the FOM dynamics in the resolved coarse space. The right-hand side of~\eqref{eq:Mori-Zwanzig-equation2}, referred to as the \textit{memory} term, due to its integration involving the past states~\cite[Chapter~6.4]{Chorin2009}, serves as a closure model accounting for the unresolved scales. Given the computational challenges associated with evaluating the memory term directly, practical approximations are commonly used. One such approach, as suggested in~\cite{Parish2017}, involves the use of the $\tau$-Model
\begin{align}\label{eq:tau_model}
	\int_0^t e^{(t-s)\bm{\mathcal{L}}} \tilde{\bm{\Pi}}\bm{\mathcal{L}} e^{s\bar{\bm{\Pi}}\bm{\mathcal{L}}} \bar{\bm{\Pi}}\bm{\mathcal{L}}\tilde{\bm{a}}_0ds \approx \tau \tilde{\bm{\Phi}}^{T} \bm{J}(\tilde{\bm{u}}(t)) \left[\bar{\bm{\Pi}} \bm{R}(\tilde{\bm{u}}(t))\right],
\end{align}
where $\bm{J}(\tilde{\bm{u}}(t)) = \frac{\partial \bm{R}(\tilde{\bm{u}}(t))}{\partial \tilde{\bm{u}}}$ represents the Jacobian of the right-hand side operator in the resolved coarse space. The scalar $\tau \in \mathbb{R}^+$ assumes the role of a \textit{memory length}, serving as a tuning parameter to enhance accuracy and stability of the APG-ROM~\cite{Parish2020}.

\subsection{Summary of the APG-ROM}
The steps taken in Sections~\ref{sec:Liouville-Equation} and~\ref{sec:Mori-Zwanzig-Equation} lead to the general APG-ROM formulation
\begin{align}\label{eq:APG_ROM_general}
	\tilde{\bm{\Phi}}^{T}\left( \dot{\tilde{\bm{u}}}^*(t) - \bm{R}(\tilde{\bm{u}}(t))\right) = \tau \tilde{\bm{\Phi}}^{T} \bm{J}(\tilde{\bm{u}}(t)) \left[\bar{\bm{\Pi}} \bm{R}(\tilde{\bm{u}}(t))\right].
\end{align}
The left-hand side of~\eqref{eq:APG_ROM_general} shows the standard G-ROM, while the right-hand side introduces an additional closure model.
In the Petrov-Galerkin form, where trial and test bases differ, the authors in \cite{Parish2020} reformulated the expression~\eqref{eq:APG_ROM_general} to $\left(\left(\bm{I} + \tau \bar{\bm{\Pi}}^T \bm{J}^T(\tilde{\bm{u}}(t)) \right)\tilde{\bm{\Phi}}\right)^T \left(\dot{\tilde{\bm{u}}}^*(t) - \bm{R}(\tilde{\bm{u}}(t))\right)=0$.
This transformation leads to the definition of the test basis
\begin{align}\label{eq:APG_Modes}
	\bm{\Psi}^{\text{APG}}(\tilde{\bm{u}}(t)) = \left(\left(\bm{I} + \tau \bar{\bm{\Pi}}^T \bm{J}^T(\tilde{\bm{u}}(t)) \right)\tilde{\bm{\Phi}}\right)^T.
\end{align}
We stress that $\bm{\Psi}^{\text{APG}}$ in \eqref{eq:APG_Modes}, in contrast to $\bm{\Phi}$ and $\bm{\Psi}$, is not constant but a function of time. This implies $\bm{\Psi}^{\text{APG}}$ needs to be computed in every time step. Moreover, this evaluation occurs in the spatial domain $\mathbb{R}^N$, and as $N$ increases, the computational benefits inherent in ROMs are compromised.

\section{Efficient Adjoint Petrov-Galerkin Reduced Order Model for fluid flows}\label{sec:APG_ROM_NSE}
To address the issues associated with the APG-ROM, we exploit the polynomial structure of the incompressible Navier-Stokes equations. This leads to the main contribution of the paper, the eAPG-ROM. In contrast to $\bm{\Psi}^{\text{APG}}$ in \eqref{eq:APG_Modes}, the test basis need not be evaluated at each time step; rather, the structure of the APG-ROM is adjusted accordingly. Furthermore, by exploiting the polynomial nature of the Navier-Stokes equations, the online evaluation no longer takes place in $\mathbb{R}^N$ but rather in the more manageable order of $\mathbb{R}^{r^3}$, where $r$ represents the dimension of the ROM. In addition, we separate the offline phase and online phase of the APG-ROM, relocating the high computational costs to the offline phase. This will result in a streamlined APG-ROM, larger than the G-ROM yet computationally more efficient than the general APG formulation.

Substituting $\dot{\tilde{\bm{u}}} = \tilde{\bm{\Phi}} \dot{\tilde{\bm{a}}}(t)$ and using $\tilde{\bm{\Phi}}^T \tilde{\bm{\Phi}} = \bm{I}$ in \eqref{eq:APG_ROM_general} yields
\begin{alignat}{4}\label{eq:APG_ROM_general_rearranged}
			&&\dot{\tilde{\bm{a}}}(t) &&= \tilde{\bm{\Phi}}^{T} \left(\bm{R}(\tilde{\bm{u}}(t)) + \tau \bm{J}(\tilde{\bm{u}}(t)) \left[\bar{\bm{\Pi}} \bm{R}(\tilde{\bm{u}}(t))\right]\right).
\end{alignat}
We derive the exact Jacobian operator $\bm{J}(\tilde{\bm{u}}(t))[\bm{v}]$ with respect to an arbitrary vector valued function $\bm{v}$, resulting in
\begin{align}\label{eq:Jacobian1}
	\bm{J}(\tilde{\bm{u}}(t))[\bm{v}] 
	&= \frac{\partial \bm{R}(\tilde{\bm{u}}(t))}{\partial \tilde{\bm{u}}}[\bm{v}]\nonumber\\
	&= \frac{\partial}{\partial \tilde{\bm{u}}}\big(-(\tilde{\bm{u}}(t)\cdot \bm{\nabla})\tilde{\bm{u}}(t) +\nu\bm{\Delta} \tilde{\bm{u}}(t)\big)[\bm{v}]\nonumber\\
	&= -\bm{\nabla} \tilde{\bm{u}}(t)[\bm{v}] - (\tilde{\bm{u}}(t)\cdot\bm{\nabla})[\bm{v}] + \nu\bm{\Delta}[\bm{v}].
\end{align}
Substituting $\tilde{\bm{u}}(t)$ with the resolved coarse-scale basis vectors and modal coefficients on the right-hand side of~\eqref{eq:Jacobian1} now yields
\begin{align}\label{eq:jacobi_NSE}
	\bm{J}(\tilde{\bm{u}}(t))[\bm{v}]
	&= 
		-\bm{\nabla} \left(\bm{u}' + \tilde{\bm{\Phi}} \tilde{\bm{a}}(t)\right)[\bm{v}]
		- \left(\left(\bm{u}' + \tilde{\bm{\Phi}} \tilde{\bm{a}}(t)\right)\cdot\bm{\nabla}\right) [\bm{v}]
		+ \nu\bm{\Delta}[\bm{v}]\nonumber\\
	&= 
		- \bm{\nabla} \tilde{\bm{\Phi}} \tilde{\bm{a}}(t) [\bm{v}]
		- \left(\tilde{\bm{\Phi}} \tilde{\bm{a}}(t)\cdot\bm{\nabla}\right)  [\bm{v}]
		- \bm{\nabla} \bm{u}'[\bm{v}]
		- \left(\bm{u}'\cdot\bm{\nabla}\right)  [\bm{v}]
		+ \nu\bm{\Delta}[\bm{v}].
\end{align}
The subsequent step involves applying the fine-scale projection operator $\bar{\bm{\Pi}}$ to the right-hand side operator $\bm{R}(\tilde{\bm{u}}(t))$ in \eqref{eq:G_ROM_NS_Arranged_notProjected}, where
\begin{align}\label{eq:RHS_orthgonalProjected}
	\bar{\bm{\Pi}} \bm{R}(\tilde{\bm{u}}(t)) 
	&= 
		\left(\bm{I} - \tilde{\bm{\Phi}}\tilde{\bm{\Phi}}^{T}\right)
		\Big(	\bm{Q}^{\text{G},N}  \big[\tilde{\bm{a}}(t)\otimes \tilde{\bm{a}}(t)\big] 
				+ \bm{L}^{\text{G},N}  \tilde{\bm{a}}(t) 
				+ \bm{C}^{\text{G},N}
		\Big)\nonumber\\
	&=
		\left(
			\bm{I} - \tilde{\bm{\Phi}}\tilde{\bm{\Phi}}^{T}
		\right)
		\bm{Q}^{\text{G},N}  \big[\tilde{\bm{a}}(t)\otimes \tilde{\bm{a}}(t)\big]
		+ \left(
			\bm{I} - \tilde{\bm{\Phi}}\tilde{\bm{\Phi}}^{T}
		\right)
		\bm{L}^{\text{G},N}  \tilde{\bm{a}}(t)
		+ \left(
			\bm{I} - \tilde{\bm{\Phi}}\tilde{\bm{\Phi}}^{T}
		\right)
		\bm{C}^{\text{G},N}\nonumber\\
	&=
		\bm{Q}^{\bar{\bm{\Pi}}}  \big[\tilde{\bm{a}}(t)\otimes \tilde{\bm{a}}(t)\big]
		+ \bm{L}^{\bar{\bm{\Pi}}}  \tilde{\bm{a}}(t)
		+ \bm{C}^{\bar{\bm{\Pi}}}.
\end{align}
The Jacobian operator~\eqref{eq:jacobi_NSE} applied to~\eqref{eq:RHS_orthgonalProjected} results in
\begin{align*}
		\bm{J}(\tilde{\bm{u}}(t)) \left[\bar{\bm{\Pi}} \bm{R}(\tilde{\bm{u}}(t))\right]
	=&
		\left(
			- \bm{\nabla} \tilde{\bm{\Phi}} \tilde{\bm{a}}(t) 
			- \tilde{\bm{\Phi}} \tilde{\bm{a}}(t)\cdot\bm{\nabla} 
			- \bm{\nabla} \bm{u}'
			- \bm{u}'\cdot\bm{\nabla} 
			+ \nu\bm{\Delta}
		\right)\\
		&\Big(
			\bm{Q}^{\bar{\bm{\Pi}}}  \big[\tilde{\bm{a}}(t)\otimes \tilde{\bm{a}}(t)\big]
			+ \bm{L}^{\bar{\bm{\Pi}}}  \tilde{\bm{a}}(t)
			+ \bm{C}^{\bar{\bm{\Pi}}}
		\Big).
\end{align*}
Rearranging the space-dependent parts yields
\begin{align}\label{eq:APG_ROM_Jacobian_onto_orthogonalizedRHS}
	\begin{split}
		\bm{J}(\tilde{\bm{u}}(t)) \left[\bar{\bm{\Pi}} \bm{R}(\tilde{\bm{u}}(t))\right]
		=
		\bm{K}^{\text{eAPG},N}  \big[\tilde{\bm{a}}(t)\otimes \tilde{\bm{a}}(t)\otimes \tilde{\bm{a}}(t)\big]
		+ \bm{Q}^{\text{eAPG},N}  \big[\tilde{\bm{a}}(t)\otimes \tilde{\bm{a}}(t)\big]
		+\bm{L}^{\text{eAPG},N}  \tilde{\bm{a}}(t)
		+ \bm{C}^{\text{eAPG},N},
	\end{split}
\end{align}
where
\begin{equation}\label{eq:APG_ROM_coeffs_spatial}
	\begin{aligned}
		\bm{K}^{\text{eAPG},N} &= 	- \bm{\nabla} \tilde{\bm{\Phi}} \bm{Q}^{\bar{\bm{\Pi}}}
									- (\tilde{\bm{\Phi}} \cdot\bm{\nabla}) \bm{Q}^{\bar{\bm{\Pi}}}
									\in\mathbb{R}^{N\times r^3},\\
		\bm{Q}^{\text{eAPG},N} &= 	- \bm{\nabla} \tilde{\bm{\Phi}} \bm{L}^{\bar{\bm{\Pi}}}
									- (\tilde{\bm{\Phi}} \cdot\bm{\nabla}) \bm{L}^{\bar{\bm{\Pi}}}
									- \bm{\nabla} \bm{u}' \bm{Q}^{\bar{\bm{\Pi}}}
									- (\bm{u}'\cdot\bm{\nabla}) \bm{Q}^{\bar{\bm{\Pi}}}
									+ \nu\bm{\Delta} \bm{Q}^{\bar{\bm{\Pi}}}
									\in\mathbb{R}^{N\times r^2},\\
		\bm{L}^{\text{eAPG},N} &= 	- \bm{\nabla} \tilde{\bm{\Phi}} \bm{C}^{\bar{\bm{\Pi}}}
									- (\tilde{\bm{\Phi}} \cdot\bm{\nabla}) \bm{C}^{\bar{\bm{\Pi}}}
									- \bm{\nabla} \bm{u}' \bm{L}^{\bar{\bm{\Pi}}}
									- (\bm{u}'\cdot\bm{\nabla}) \bm{L}^{\bar{\bm{\Pi}}}
									+ \nu\bm{\Delta} \bm{L}^{\bar{\bm{\Pi}}}
							\in\mathbb{R}^{N\times r},\\
		\bm{C}^{\text{eAPG},N} &= 	- \bm{\nabla} \bm{u}' \bm{C}^{\bar{\bm{\Pi}}}
									- (\bm{u}'\cdot\bm{\nabla}) \bm{C}^{\bar{\bm{\Pi}}}
									+ \nu\bm{\Delta} \bm{C}^{\bar{\bm{\Pi}}}
									\in\mathbb{R}^{N}.\\
	\end{aligned}
\end{equation}
The final step involves assembling the right-hand side of~\eqref{eq:APG_ROM_Jacobian_onto_orthogonalizedRHS} and projecting it to the coarse-scale trial basis $\tilde{\bm{\Phi}}$ as in~\eqref{eq:APG_ROM_general}. This process yields the time-invariant eAPG-ROM for the incompressible Navier-Stokes equations
\begin{subequations}\label{eq:APG_ROM}
	\begin{align}
		\begin{split}\label{eq:APG_ROM_eq}
			\dot{\tilde{\bm{a}}}^{\text{eAPG}}(t)
			=\,
					&\bm{K}^{\text{eAPG}}  \big[\tilde{\bm{a}}^{\text{eAPG}}(t)\otimes \tilde{\bm{a}}^{\text{eAPG}}(t)\otimes \tilde{\bm{a}}^{\text{eAPG}}(t)\big]\\
					&+ \bm{Q}^{\text{eAPG}}  \big[\tilde{\bm{a}}^{\text{eAPG}}(t)\otimes \tilde{\bm{a}}^{\text{eAPG}}(t)\big]\\
					&+ \bm{L}^{\text{eAPG}}  \tilde{\bm{a}}^{\text{eAPG}}(t)\\
					&+ \bm{C}^{\text{eAPG}}
		\end{split}\\
		\begin{split}\label{eq:APG_ROM_initialCondition}
			\tilde{\bm{a}}^{\text{eAPG}}(0) =&\, \tilde{\bm{a}}_0 =\, \tilde{\bm{\Phi}}^T \bm{u}^*_0,
		\end{split}
	\end{align}
\end{subequations}
where
\begin{equation}\label{eq:APG_ROM_coeffs}
	\begin{aligned}
		\bm{K}^{\text{eAPG}} &= \tau \tilde{\bm{\Phi}}^T \bm{K}^{\text{eAPG},N}\in\mathbb{R}^{r\times r^3},\\
		\bm{Q}^{\text{eAPG}} &= \tilde{\bm{\Phi}}^T \bm{Q}^{\text{G},N} + \tau \tilde{\bm{\Phi}}^T\bm{Q}^{\text{eAPG},N}\in\mathbb{R}^{r\times r^2},\\
		\bm{L}^{\text{eAPG}} &= \tilde{\bm{\Phi}}^T \bm{L}^{\text{G},N} + \tau \tilde{\bm{\Phi}}^T\bm{L}^{\text{eAPG},N}\in\mathbb{R}^{r\times r},\\
		\bm{C}^{\text{eAPG}} &= \tilde{\bm{\Phi}}^T \bm{C}^{\text{G},N} + \tau \tilde{\bm{\Phi}}^T\bm{C}^{\text{eAPG},N}\in\mathbb{R}^{r}.
	\end{aligned}
\end{equation}
The eAPG-ROM~\eqref{eq:APG_ROM_eq} can be numerically integrated for $\tilde{\bm{a}}^\text{eAPG}(t)$ with the specified initial condition~\eqref{eq:APG_ROM_initialCondition} and its coefficients~\eqref{eq:APG_ROM_coeffs}. Subsequently, the velocity field is reconstructed using~\eqref{eq:FOM_exact_approx_in_R}.
The right-hand side of \eqref{eq:APG_ROM_eq} is cubic, in contrast to the quadratic nature of the G-ROM, and introduces additional polynomial terms. This however, only marginally increases in the computational cost of the online phase compared to the G-ROM (see Section~\ref{sec:Implementation_and_computationalCost}). The coefficients~\eqref{eq:APG_ROM_coeffs_spatial} of the eAPG-ROM are detailed in~\ref{ap:Matrices_APG_ROM}.

\subsection{eAPG-ROM tuning: Determining the memory length}\label{sec:APG-ROM_opt}
Previous studies showed a correlation between the memory length $\tau$ and the eigenvalues of the Jacobian of the right-hand side operator $\bm{J}(\tilde{\bm{u}}(t))$ for linear systems~\cite{Parish2017, Parish2020}. However, an extension to nonlinear systems introduces significant challenges, and the adaptation to such systems remains unclear~\cite{Parish2017}. Consequently, a heuristic approach, as detailed in~\cite{Parish2017}, was adopted, which yields
\begin{equation}\label{eq:memoryLength}
	\begin{aligned}
		\tau 
		&= w\left(\rho\left(\tilde{\bm{\Phi}}^T \bm{J}(\tilde{\bm{u}}(t)) [\tilde{\bm{\Phi}}]\right)\right)^{-1}\\
		&= w\left(\rho\left(\tilde{\bm{\Phi}}^T \left(-\nabla\tilde{\bm{u}}(t)[\tilde{\bm{\Phi}}] -(\tilde{\bm{u}}(t)\cdot\nabla)[\tilde{\bm{\Phi}}] + \nu\Delta [\tilde{\bm{\Phi}}]\right)\right)\right)^{-1},
	\end{aligned}
\end{equation}
where $\rho: \mathbb{R}^r \rightarrow \mathbb{R}$ and $w\in\mathbb{R}^+$ are the spectral radius and a weighting parameter, respectively. The approach involves computing the memory length by evaluating the Jacobian $\bm{J}(\tilde{\bm{u}}(t))[\tilde{\bm{\Phi}}]$ only for the first time step and keeping it constant throughout the simulation. An alternative approach involves introducing a time-dependent memory length, updating $\tau$ over time, using the same heuristic \cite{Parish2020}. However,~\cite{Parish2020} showed that this proves to be useful only if $\bm{J}(\tilde{\bm{u}}(t))[\tilde{\bm{\Phi}}]$ changes drastically over time. During our numerical analysis, we did not observe significant temporal changes in the Jacobian; consequently, the adaptive method was not implemented. 

Also in \cite{Parish2020}, the authors proposed a method for determining an optimal memory length by minimizing the error between ROM and FOM at only a single time step. Extending this minimization over the complete time period would, presumably, have been numerically prohibitive, as the general APG-ROM is significantly more computationally intensive than the eAPG-ROM (see Section~\ref{sec:FLOPS}). Consequently, due to the efficient formulation of the eAPG-ROM, we are now able to apply and propose an empirical approach that uses numerical optimization over the full time period.

This optimization aims to minimize the error between modal coefficients of the eAPG-ROM $\tilde{\bm{a}}^{\text{eAPG}}(t_m)$ and modal coefficients obtained from projecting the coarse-scale trial basis onto the time variant contributions of the velocities from the sampled FOM solution $\tilde{\bm{a}}^{\text{POD}}(t_m) = \tilde{\bm{\Phi}}^T \bm{u}^*(t_m)$, i.e., 
\begin{equation}\label{eq:memoryLength_opt}
	\begin{aligned}
	&\minA_{w} \sum_{m=1}^M \lVert \tilde{\bm{a}}^{\text{eAPG}}(t_m)-\tilde{\bm{a}}^{\text{POD}}(t_m))\rVert^2_2,
	\end{aligned}
\end{equation}
with initial value $w=1$.
Note that although the ROMs are initially formulated as continuous in time, they are typically solved using numerical integration schemes, resulting in sampled solutions. 
Solving~\eqref{eq:memoryLength_opt} involves evaluating the eAPG-ROM~\eqref{eq:APG_ROM} and recalculating the memory length~\eqref{eq:memoryLength} while keeping the Jacobian $\bm{J}(\tilde{\bm{u}}(t))[\tilde{\bm{\Phi}}]$ constant in every iteration of the optimization algorithm. 
The weight obtained with~\eqref{eq:memoryLength_opt} is used for computing the optimal memory length with~\eqref{eq:memoryLength}, denoted as $\tau^{\text{opt}}$. This optimized memory length is subsequently applied to the eAPG-ROMs coefficients~\eqref{eq:APG_ROM_coeffs}. To distinguish the modal coefficients obtained from the eAPG-ROM with the optimal memory length $\tau^{\text{opt}}$ from those associated with an arbitrarily chosen memory length $\tau$, we denote the former as $\tilde{\bm{a}}^{\text{eAPG},\tau,{\text{opt}}}(t)$ and the latter $\tilde{\bm{a}}^{\text{eAPG},\tau}(t)$.

\subsection{Memory length augmentation}\label{sec:MemoryTerm_matrix}
We demonstrate the efficacy of the proposed approach in Section~\ref{sec:numerical_example}. However, the obtained results do not align perfectly with those of the FOM. This leads to the second key contribution of this paper: a refinement to the memory length approach. Until now, the memory length has been treated as a scalar representing a single timescale. We augment the memory length by incorporating supplementary timescales, effectively expanding the scope of $\tau$ as shown in~\eqref{eq:tau_model}. This is realized by a matrix memory length $\bm{T} \in\mathbb{R}^{r\times r}$. The equation for $\bm{T}$ reads 
\begin{equation}\label{eq:memoryLength_r}
	\begin{aligned}
		\bm{T}
		&= \bm{W}\left(\rho\left(\tilde{\bm{\Phi}}^T \bm{J}(\tilde{\bm{u}}(t)) [\tilde{\bm{\Phi}}]\right)\right)^{-1}\\
		&= \bm{W}\left(\rho\left(\tilde{\bm{\Phi}}^T \left(-\nabla\tilde{\bm{u}}(t)[\tilde{\bm{\Phi}}] -(\tilde{\bm{u}}(t)\cdot\nabla)[\tilde{\bm{\Phi}}] + \nu\Delta \nabla[\tilde{\bm{\Phi}}]\right)\right)\right)^{-1}.
	\end{aligned}
\end{equation}
The memory term requires positive time scales, which implies that the matrix $\bm{T}$ must be positive definite. Note that constraining $\bm{T}$ to be diagonal is overly restrictive. Numerical investigations have shown that the modal coefficients are not decoupled in the memory integral. We also extend the weight to be a positive definite weighting matrix $\bm{W}\in\mathbb{R}^{r\times r}$. 
% The extension of the memory length to a matrix, along with the associated weighting matrix, introduces additional degrees of freedom in the optimization process and can also be regarded as a strategy to augment the number of tuning parameters. 
% In conclusion, this increased flexibility allows for a finer adjustment of the eAPG-ROM's dynamics and 
Our analysis in Section~\ref{sec:numerical_example} demonstrates that the transition from $\tau\in\mathbb{R}^+$ to $\bm{T} \in\mathbb{R}^{r\times r}$ contributes significantly to the accuracy and stability of the eAPG-ROM.
The optimization~\eqref{eq:memoryLength_opt} now reads 
\begin{equation}\label{eq:memoryLength_r_opt}
	\begin{aligned}
	&\minA_{\bm{W}} \sum_{m=1}^M \lVert \tilde{\bm{a}}^{\text{eAPG}}(t_m)-\tilde{\bm{a}}^{\text{POD}}(t_m)\rVert^2_2,
	\end{aligned}
\end{equation}
with the initial value $\bm{W}=\bm{I}$. Solving~\eqref{eq:memoryLength_r_opt} involves an iterative procedure. In each iteration, the eAPG-ROM~\eqref{eq:APG_ROM} is evaluated while the memory length is recalculated based on~\eqref{eq:memoryLength_r}, with the Jacobian $\bm{J}(\tilde{\bm{u}}(t))[\tilde{\bm{\Phi}}]$ held constant throughout the optimization algorithm.
The resulting optimal weighting matrix $\bm{T}^{\text{opt}}$ is subsequently applied in the eAPG-ROMs coefficients~\eqref{eq:APG_ROM_coeffs}. We denote the modal coefficients obtained with the optimal matrix memory length $\bm{T}^{\text{opt}}$ as $\tilde{\bm{a}}^{\text{eAPG},\bm{T},{\text{opt}}}(t)$. Note that the results obtained with $W = w\bm{I}$ are equivalent to those achieved with the scalar memory length.

Numerical investigations show that optimizing over a single flow period may not always yield the best results. Given the periodic nature of the flow under consideration, extending the optimization process to include multiple periods typically improves the results. Accordingly, this approach will be used in Section~\ref{sec:ROM_Karman}.

\section{Error estimation}\label{sec:Error_Eval}
We use a 2-norm error to evaluate the precision of the ROMs derived in Section~\ref{sec:G-ROM} and~\ref{sec:APG_ROM_NSE}. First, the squared 2-norm is employed to quantify the difference between the sampled velocity snapshots from the FOM $\bm{u}^*(t_m)$ and those obtained through projection onto the coarse-scale basis $\tilde{\bm{u}}^*(t_m)$. This discrepancy equals the sum of singular values neglected during the truncation process (see, e.g.,~\cite[Chapter~2.4.2]{Golub2013})
\begin{align*}
	\sum_{m=1}^M\lVert\big( \bm{u}^*(t_m) - \tilde{\bm{u}}^*(t_m)\big)\rVert_2^2 
	= 
	\sum_{k=r+1}^M\sigma_k^2.
\end{align*}
This represents a lower bound for the error in both ROMs. The G-ROM and eAPG-ROM contribute to an additional error that depends on the modal coefficients (see~\cite[Appendix C]{Sommer2023})
\begin{align*}
	\sum_{m=1}^M\lVert\big( \bm{u}^*(t_m) - \tilde{\bm{\Phi}}\tilde{\bm{a}}^{\text{ROM}}(t_m)\big)\rVert_2^2
	= 
	\sum_{k=r+1}^M\sigma_k^2 + \sum_{m=1}^M\lVert\tilde{\bm{a}}^{\text{POD}}(t_m) - \tilde{\bm{a}}^{\text{ROM}}(t_m)\rVert_2^2,
\end{align*}
where $\tilde{\bm{a}}^{\text{ROM}}$ represents either the sampled solution of the G-ROM $\tilde{\bm{a}}^{\text{G}}$ from \eqref{eq:G_ROM} or the eAPG-ROM $\tilde{\bm{a}}^{\text{eAPG}}$ from \eqref{eq:APG_ROM}.
To evaluate the overall error, a normalization by the sum of singular values is performed
\begin{equation*}
	\begin{aligned}
		\mathcal{E}_{\text{Total}}(r) 
		= 
		\frac{\sum_{k=r+1}^M\sigma_k^2 + \sum_{m=1}^M\lVert\tilde{\bm{a}}^{\text{POD}}(t_m) - \tilde{\bm{a}}(t_m)\rVert_2^2}{\sum_{k=1}^M\sigma_k^2}.
	\end{aligned}
\end{equation*} 
This normalization ensures consistency with the truncation error~\eqref{sec:TruncationError}. The resulting total error $\mathcal{E}_{\text{Total}}(r)$ is decomposed into the sum of the truncation error $\mathcal{E}_{\text{TRU}}(r)$ and the ROM error $\mathcal{E}_{\text{ROM}}(r)$
\begin{align*}
	\mathcal{E}_{\text{Total}}(r)=\mathcal{E}_{\text{TRU}}(r)+\mathcal{E}_{\text{ROM}}(r),
\end{align*}
where
\begin{align*}
    \mathcal{E}_{\text{ROM}}(r)=\frac{\sum_{m=1}^M\lVert\tilde{\bm{a}}^{\text{POD}}(t_m) - \tilde{\bm{a}}(t_m)\rVert_2^2}{\sum_{k=1}^M\sigma_k^2}.
\end{align*}
Furthermore, a 2-norm of the error between $\bm{u}(t_m)$ and $\tilde{\bm{u}}(t_m)$ is introduced. This figure quantifies the error between the velocities from the FOM and the coarse-scale projection, normalized by the reference value $u_{\text{ref}}$ and averaged over the grid points and time steps
\begin{align*}
	\mathcal{E}_{\text{REC}}
	=\frac{1}{dN_{\text{grid}}M}\sum_{m=1}^M \frac{||\bm{u}(t_m)-\tilde{\bm{u}}(t_m)||_2}{u_{\text{ref}}}.
\end{align*}
The reference velocity $u_{\text{ref}}$ is specified in Section~\ref{sec:numerical_example}.

\section{Implementation and computational cost}\label{sec:Implementation_and_computationalCost}
In this section, we provide algorithms for the implementation of the standard G-ROM, outlined in Section~\ref{sec:Galerkin_Projection}, the APG-ROM from Section~\ref{sec:APG_ROM_general} and the efficient variant derived in this paper, discussed in Section~\ref{sec:APG_ROM_NSE}. These algorithms are tailored for fluid flows governed by the incompressible Navier-Stokes equations. Furthermore, we analyze the computational effort of the three methods in terms of floating point operations (flops).

\subsection{Implementation and Algorithms}\label{sec:Implementation}
Algorithm~\ref{alg:G-ROM_Offline} and~\ref{alg:G-ROM_Online} detail the offline and online, respectively, calculations necessary for the G-ROM. Algorithm~\ref{alg:APG-ROM_Offline} and~\ref{alg:APG-ROM_Online} are the corresponding algorithms for the eAPG-ROM. Algorithm~\ref{alg:General-APG-ROM_Online} for the APG-ROM from Section~\ref{sec:APG_ROM_general} is included for completeness. Note that the evaluation of the right-hand-side operator for the general APG-ROM in Algorithm~\ref{alg:General-APG-ROM_Online} necessitates operations in $\mathbb{R}^N$ in each time step, and the polynomial form of the incompressible Navier-Stokes equations is not exploited.

For all three reduced models, G-ROM, eAPG-ROM and APG-ROM, integrations of the ODEs for the model coefficients $\tilde{\bm{a}}(t)$ need to be carried out (see Algorithms~\ref{alg:G-ROM_Online},~\ref{alg:APG-ROM_Online},~\ref{alg:General-APG-ROM_Online}). While we use an adaptive method, specifically the Dormand-Prince integration method (see, e.g.~\cite{Dormand1980}), for all numerical calculations in Section~\ref{sec:numerical_example}, we use Euler's explicit method in Algorithms~\ref{alg:G-ROM_Online},~\ref{alg:APG-ROM_Online},~\ref{alg:General-APG-ROM_Online}, because it is straightforward to count flops in this case. Consequently, the flop estimates derived from this basic scheme serve as a convenient baseline for comparison with more advanced schemes. However, higher-order or adaptive methods can, in certain problems, require fewer time steps, which may sometimes offset their additional per-step overhead. Therefore, the reported flop count should be regarded as a convenient benchmark rather than an absolute lower bound.

We summarize some technical details that apply in this section. The time step of Euler's method is denoted as $\Delta t$. In accordance with the assumptions explained in Section~\ref{sec:Galerkin_Projection}, pressure is neglected during the evaluation of the right-hand side operator. We use the modified matrix representation~\eqref{eq:memoryLength_r} for the memory length in Algorithms~\ref{alg:APG-ROM_Offline} and~\ref{alg:General-APG-ROM_Online}, as detailed in Section~\ref{sec:MemoryTerm_matrix}. In contrast to the finite-difference approximation used in~\cite{Parish2020}, the Jacobian used in the eAPG-ROM is exactly linearized with~\eqref{eq:jacobi_NSE}, as discussed in Section~\ref{sec:APG_ROM_NSE}. The computation of the fine-scale projection operator $\bar{\bm{\Pi}} = \bm{I} - \tilde{\bm{\Phi}}\tilde{\bm{\Phi}}^{T}$ can result in a prohibitively large memory consumption. Therefore, when calculating $\bar{\bm{\Pi}} \bm{R}(\tilde{\bm{u}}(t)) = (\bm{I} - \tilde{\bm{\Phi}}\tilde{\bm{\Phi}}^{T}) \bm{R}(\tilde{\bm{u}}(t))$, it is advisable not to precalculate $\bar{\bm{\Pi}}$ but rather split the computation into $\bar{\bm{\Pi}} \bm{R}(\tilde{\bm{u}}(t)) = \bm{R}(\tilde{\bm{u}}(t)) - \tilde{\bm{\Phi}}\tilde{\bm{\Phi}}^{T} \bm{R}(\tilde{\bm{u}}(t))$. This involves projecting $\tilde{\bm{\Phi}}^{T} \bm{R}(\tilde{\bm{u}}(t))$ and then projecting it back, resulting in $\bar{\bm{\Pi}} \bm{R}(\tilde{\bm{u}}(t)) = \bm{R}(\tilde{\bm{u}}(t)) - \tilde{\bm{\Phi}}(\tilde{\bm{\Phi}}^{T} \bm{R}(\tilde{\bm{u}}(t)))$. This approach reduces the memory demands considerably.

\begin{algorithm2e}[H]
	\SetKwBlock{Steps}{Steps:}{end}
	\caption{G-ROM for fluid flows governed by the incompressible Navier-Stokes equations: Offline phase}\label{alg:G-ROM_Offline}
	\KwData{$\tilde{\bm{\Phi}}$, $\bm{u}'$, $\nu$\;}
	\SetAlgoLined
	\Steps()
	{
		\textbf{1.}\quad Compute spatial derivatives $\dfrac{\partial \tilde{\bm{\Phi}}_i}{\partial x_j}$, $\dfrac{\partial^2 \tilde{\bm{\Phi}}_i}{\partial x_j^2}$, $\dfrac{\partial \bm{u}'_i}{\partial x_j}$, $\dfrac{\partial^2 \bm{u}'_i}{\partial x_j^2}$, $i,j=1,\hdots,d$\\
	
		\textbf{2.}\quad Compute coefficient tensors $\bm{Q}^{\text{G},N}$, $\bm{L}^{\text{G},N}$, $\bm{C}^{\text{G},N}$ in $\mathbb{R}^N$ according to~\eqref{eq:G_ROM_NS_Arranged_notProjected_coefficients} and~\ref{ap:Matrices_G_ROM}\\

		\textbf{3.}\quad Compute final coefficient tensors $\bm{Q}^{\text{G}}$, $\bm{L}^{\text{G}}$, $\bm{C}^{\text{G}}$ in $\mathbb{R}^r$  by projection according to~\eqref{eq:G_ROM_coeffs}\\
	}
\end{algorithm2e}
\begin{algorithm2e}[H]
	\SetKwBlock{loop}{loop:}{end}
	\caption{G-ROM for fluid flows governed by the incompressible Navier-Stokes equations: Online phase}\label{alg:G-ROM_Online}
	\KwData{$\tilde{\bm{a}}(t_{m})$, $\Delta t$, $\bm{Q}^{\text{G}}$, $\bm{L}^{\text{G}}$, $\bm{C}^{\text{G}}$\;}
	\SetAlgoLined
	\loop(Explicit Euler)
	{
		\textbf{1.}\quad Compute time derivative of the modal coefficients\\
		\qquad\quad$\dot{\tilde{\bm{a}}}^{\text{G}}(t_m) = \bm{Q}^{\text{G}} \big[\tilde{\bm{a}}(t_m)\otimes \tilde{\bm{a}}(t_m)\big] 
		+ \bm{L}^{\text{G}} \tilde{\bm{a}}(t_m) 
		+ \bm{C}^{\text{G}}$ according to~\eqref{eq:G_ROM_eq}\\

		\textbf{2.}\quad Update the modal coefficients\\
		\qquad\quad$\tilde{\bm{a}}^{\text{G}}(t_{m+1}) = \tilde{\bm{a}}(t_{m}) + \Delta t\dot{\tilde{\bm{a}}}^{\text{G}}(t_m)$\\
	}
\end{algorithm2e}
\begin{algorithm2e}[H]
	\SetKwBlock{Steps}{Steps:}{end}
	\caption{eAPG-ROM for fluid flows governed by the incompressible Navier-Stokes equations: Offline phase}\label{alg:APG-ROM_Offline}
	\KwData{$\tilde{\bm{\Phi}}$, $\bm{u}'$, $\nu$, $\bm{T}$\;}
	\SetAlgoLined
	\Steps()
	{
		\textbf{1.}\quad Compute spatial derivatives $\dfrac{\partial \tilde{\bm{\Phi}}_i}{\partial x_j}$, $\dfrac{\partial^2 \tilde{\bm{\Phi}}_i}{\partial x_j^2}$, $\dfrac{\partial \bm{u}'_i}{\partial x_j}$, $\dfrac{\partial^2 \bm{u}'_i}{\partial x_j^2}$, $i,j=1,\hdots,d$\\
	
		\textbf{2.}\quad Compute coefficient tensors of the G-ROM $\bm{Q}^{\text{G},N}$, $\bm{L}^{\text{G},N}$, $\bm{C}^{\text{G},N}$ in $\mathbb{R}^N$ according to~\eqref{eq:G_ROM_NS_Arranged_notProjected_coefficients} and~\ref{ap:Matrices_G_ROM}\\

		\textbf{3.}\quad Compute the fine-scale projected coefficients tensors $\bm{Q}^{\bar{\bm{\Pi}}}$, $\bm{L}^{\bar{\bm{\Pi}}}$ and $\bm{C}^{\bar{\bm{\Pi}}}$ according to~\eqref{eq:RHS_orthgonalProjected}\\
		
		\textbf{4.}\quad Compute spatial derivatives $\dfrac{\partial \bm{Q}^{\bar{\bm{\Pi}}}_i}{\partial x_j}$, $\dfrac{\partial^2 \bm{Q}^{\bar{\bm{\Pi}}}_i}{\partial x_j^2}$, $\dfrac{\partial \bm{L}^{\bar{\bm{\Pi}}}_i}{\partial x_j}$, $\dfrac{\partial^2 \bm{L}^{\bar{\bm{\Pi}}}_i}{\partial x_j^2}$, $\dfrac{\partial \bm{C}^{\bar{\bm{\Pi}}}_i}{\partial x_j}$, $\dfrac{\partial^2 \bm{C}^{\bar{\bm{\Pi}}}_i}{\partial x_j^2}$ $i,j=1,\hdots,d$\\

		\textbf{5.}\quad Compute coefficient tensors of the eAPG-ROM $\bm{K}^{\text{eAPG},N}$, $\bm{Q}^{\text{eAPG},N}$, $\bm{L}^{\text{eAPG},N}$ and $\bm{C}^{\text{eAPG},N}$ in $\mathbb{R}^N$ according to~\eqref{eq:APG_ROM_coeffs_spatial} and~\ref{ap:Matrices_APG_ROM}\\

		\textbf{6.}\quad Compute final coefficient tensors of the eAPG-ROM $\bm{K}^{\text{eAPG}}$, $\bm{Q}^{\text{eAPG}}$, $\bm{L}^{\text{eAPG}}$ and $\bm{C}^{\text{eAPG}}$ in $\mathbb{R}^r$ by projection according to~\eqref{eq:APG_ROM_coeffs}\\
	}
\end{algorithm2e}
\begin{algorithm2e}[H]
	\SetKwBlock{loop}{loop:}{end}
	\caption{eAPG-ROM for fluid flows governed by the incompressible Navier-Stokes equations: Online phase}\label{alg:APG-ROM_Online}
	\KwData{$\tilde{\bm{a}}(t_{m})$, $\Delta t$, $\bm{K}^{\text{eAPG}}$, $\bm{Q}^{\text{eAPG}}$, $\bm{L}^{\text{eAPG}}$, $\bm{C}^{\text{eAPG}}$\;}
	\SetAlgoLined
	\loop(Explicit Euler)
	{
		\textbf{1.}\quad Compute time derivative of the modal coefficients\\
		\qquad\quad$\dot{\tilde{\bm{a}}}^{\text{eAPG}}(t_m) = \bm{K}^{\text{eAPG}}  \big[\tilde{\bm{a}}(t_m)\otimes \tilde{\bm{a}}(t_m)\otimes \tilde{\bm{a}}(t_m)\big]+$\\
		\qquad\qquad\qquad\qquad\qquad
		$\bm{Q}^{\text{eAPG}} \big[\tilde{\bm{a}}(t_m)\otimes \tilde{\bm{a}}(t_m)\big]
		+ \bm{L}^{\text{eAPG}}  \tilde{\bm{a}}(t_m)
		+ \bm{C}^{\text{eAPG}}$ according to~\eqref{eq:APG_ROM_eq}\\

		\textbf{2.}\quad Update the modal coefficients\\
		\qquad\quad$\tilde{\bm{a}}^{\text{eAPG}}(t_{m+1}) = \tilde{\bm{a}}(t_{m}) + \Delta t\dot{\tilde{\bm{a}}}^{\text{eAPG}}(t_m)$\\
	}
\end{algorithm2e}
\begin{algorithm2e}[H]
	\SetKwBlock{loop}{loop:}{end}
	\caption{General APG-ROM for fluid flows governed by the incompressible Navier-Stokes equations}\label{alg:General-APG-ROM_Online}
	\KwData{$\tilde{\bm{a}}(t_{m})$, $\Delta t$, $\tilde{\bm{\Phi}}$, $\bm{u}'$, $\nu$, $\bm{T}$\;}
	\SetAlgoLined
	\loop(Explicit Euler)
	{
		\textbf{1.}\quad Compute the resolved coarse-scale state using the coarse-scale modal coefficients\\
		\qquad\quad $\tilde{\bm{u}}(t_m)=\bm{u}' + \tilde{\bm{\Phi}} \tilde{\bm{a}}(t_m)$\\

		\textbf{2.}\quad Compute spatial derivatives $\dfrac{\partial \tilde{\bm{u}}_i(t_m)}{\partial x_j}$, $\dfrac{\partial^2 \tilde{\bm{u}}_i(t_m)}{\partial x_j^2}$, $i,j=1,\hdots,d$\\

		\textbf{3.}\quad Compute the right-hand side operator $\bm{R}(\tilde{\bm{u}}(t_m)) = -(\tilde{\bm{u}}(t_m)\cdot \bm{\nabla})\tilde{\bm{u}}(t_m) +\nu\bm{\Delta} \tilde{\bm{u}}(t_m)$\\

		\textbf{4.}\quad Compute the fine-scale projected right-hand side operator $\bar{\bm{\Pi}} \bm{R}(\tilde{\bm{u}}(t_m))$\\

		\textbf{5.}\quad Compute spatial derivatives $\dfrac{\partial \bar{\bm{\Pi}} \bm{R}(\tilde{\bm{u}}_i(t_m))}{\partial x_j}$, $\dfrac{\partial^2 \bar{\bm{\Pi}} \bm{R}(\tilde{\bm{u}}_i(t_m))}{\partial x_j^2}$, $i,j=1,\hdots,d$\\

		\textbf{6.}\quad Compute the exactly linearized and fine-scale projected right-hand side operator \\
		\qquad\quad $\bm{J}(\tilde{\bm{u}}(t_m)) \left[\bar{\bm{\Pi}} \bm{R}(\tilde{\bm{u}}(t_m))\right] = \dfrac{\partial \bm{R}(\tilde{\bm{u}}(t_m))}{\partial \tilde{\bm{u}}} \left[\bar{\bm{\Pi}} \bm{R}(\tilde{\bm{u}}(t_m))\right]$\\
		\qquad\qquad$=-\bm{\nabla} \tilde{\bm{u}}(t_m)  \bar{\bm{\Pi}} \bm{R}(\tilde{\bm{u}}(t_m)) - (\tilde{\bm{u}}(t_m)\cdot\bm{\nabla}) \bar{\bm{\Pi}} \bm{R}(\tilde{\bm{u}}(t_m)) + \nu\bm{\Delta} \bar{\bm{\Pi}} \bm{R}(\tilde{\bm{u}}(t_m))$

		\textbf{7.}\quad Compute the full right-hand side of~\eqref{eq:APG_ROM_general_rearranged}\\
		\qquad\quad $\dot{\tilde{\bm{a}}}^{\text{APG}}(t_m) = \tilde{\bm{\Phi}}^{T} \bm{R}(\tilde{\bm{u}}(t_m)) + \bm{T} \tilde{\bm{\Phi}}^{T} \bm{J}(\tilde{\bm{u}}(t_m)) \left[\bar{\bm{\Pi}} \bm{R}(\tilde{\bm{u}}(t_m))\right]$\\

		\textbf{8.}\quad Update the modal coefficients\\
		\qquad\quad$\tilde{\bm{a}}^{\text{APG}}(t_{m+1}) = \tilde{\bm{a}}(t_{m}) + \Delta t \dot{\tilde{\bm{a}}}^{\text{APG}}(t_m)$\\
	}
\end{algorithm2e}

\subsection{Approximated computational cost}\label{sec:FLOPS}
We count flops of the algorithms from Section~\ref{sec:Implementation} to assess their computational complexity. We assume a scalar addition, subtraction, and multiplication each cost 1 flop~\cite{Addison1993, Ueberhuber1997}.
Since all ROMs have the computation of trial bases with POD in common, we do not include their computational effort in our comparison. 
The Jacobian and diagonal Hessian of a function $\bm{f}:\mathbb{R}^{N}\rightarrow\mathbb{R}^{N}$ are assumed to require $\omega_1 N$ and $\omega_2 N$ flops, where $\omega_1$ and $\omega_2$ are given in Section~\ref{sec:costs_Karman}. A linear in $N$ number of flops is assumed for the Jacobian and diagonal Hessian because the derivatives are approximated with finite differences.

Table~\ref{tab:Flops_G-ROM_Offline} and~\ref{tab:Flops_APG-ROM_Offline} show the flop count for the offline phases of the G-ROM and the eAPG-ROM calculated with Algorithm~\ref{alg:G-ROM_Offline} and Algorithm~\ref{alg:APG-ROM_Offline}, respectively. For the construction of the eAPG-ROM, we assume the memory length to be in the matrix version~\eqref{eq:memoryLength_r}. The computational cost is larger for the matrix memory length~\eqref{eq:memoryLength_r} than for the scalar variant~\eqref{eq:memoryLength}. The difference is however negligible, as a scalar quantity is replaced by an $r\times r$ matrix, modifying the multiplication operation accordingly. This occurs four times during the offline phase, with no impact on the online performance. Table~\ref{tab:Flops_G-ROM_and_APG-ROM_Online} presents the flop count for the online phases for the G-ROM and the eAPG-ROM from Algorithm~\ref{alg:G-ROM_Online} and Algorithm~\ref{alg:APG-ROM_Online}. Finally, Table~\ref{tab:Flops_General-APG-ROM} lists the flop counts for the general APG-ROM from Algorithm~\ref{alg:General-APG-ROM_Online}.
\begin{table}[b!]
	\centering 
	\caption{Flops required for the computation of the coefficients for the G-ROM~\eqref{eq:G_ROM_coeffs} with Algorithm~\ref{alg:G-ROM_Offline}.}
	\label{tab:Flops_G-ROM_Offline}
	\begin{tabular}{ll}\hline
		Step in Algorithm~\ref{alg:G-ROM_Offline} & Flop Count \\ \hline
		1           & $r\omega_1 N + \omega_1 N + r\omega_2 N + \omega_2 N$        \\
		2           & $2r^2dN + 5rdN + 2rN + 3dN + N$        \\
		3           & $2r^3N + 2r^2N + 2rN - r^3 - r^2 - r$ \\ \hline
		\multirow{2}{*}{Total: $flops_{\text{G,Offline}}=$}	& $\big(2r^3 + (2d + 2)r^2 + (5d + \omega_1 + \omega_2 + 4)r + 3d + \omega_1 + \omega_2 + 1\big)N$\\
								& \quad$- r^3 - r^2 - r$ \\\hline
	\end{tabular}
	\bigskip
\end{table}
\begin{table}[b!]
	\centering 
	\caption{Flops required for the computation of the coefficients for the eAPG-ROM~\eqref{eq:APG_ROM_coeffs} with Algorithm~\ref{alg:APG-ROM_Offline}.}
	\label{tab:Flops_APG-ROM_Offline}
	\begin{tabular}{ll}\hline
		Step in Algorithm~\ref{alg:APG-ROM_Offline} & Flop Count \\ \hline
		1           & $r\omega_1 N + \omega_1 N + r\omega_2 N + \omega_2 N$ \\
		2           & $2r^2dN + 5rdN + 2rN + 3dN + N$ \\
		3           & $4r^3N + 4r^2N + 4rN - r^3 - r^2 - r$ \\ 
		4           & $r^2\omega_1N + r^2\omega_2N + r\omega_1N + r\omega_2N + \omega_1N + \omega_2N$ \\
		5           & $4r^3dN + r^3N + 9r^2dN + 4r^2N + 9rdN + 4rN + 5dN + 2N$ \\
		6           & $2r^4N + 2r^3N + 2r^2N + 2rN - r^4 - r^3 - r^2 - r$ \\ \hline
		\multirow{2}{*}{Total: $flops_{\text{eAPG,Offline}}=$} 	& $\big(2r^4 + (4d + 7)r^3 + (11d + w1 + w2 + 10)r^2 + (14d + 2w1 + 2w2 + 12)r$ \\
								& \quad$+ 8d + 2w1 + 2w2 + 3\big)N - r^4 - 2r^3 - 2r^2 - 2r$ \\\hline
	\end{tabular}
	\bigskip
\end{table}
\begin{table}[b!]
	\centering 
	\caption{Flops required for the online evaluation of the G-ROM~\eqref{eq:G_ROM_eq} and eAPG-ROM~\eqref{eq:APG_ROM_eq} for a single time step using an explicit Euler integration and Algorithm~\ref{alg:G-ROM_Online} and ~\ref{alg:APG-ROM_Online}.}
	\label{tab:Flops_G-ROM_and_APG-ROM_Online}
	\begin{tabular}{lll}\hline
		Method  & Step & Flop Count \\ \hline
		\multirow{3}{*}{G-ROM online phase in Algorithm~\ref{alg:G-ROM_Online}}   & 1    & $2r^3 + 2r^2 + r$      \\
			& 2    	& $2r$      			\\\cline{2-3} 
			& Total: $flops_{\text{G,Online}}=$	& $2r^3 + 2r^2 + 3r$	\\
		\hline
		\multirow{3}{*}{eAPG-ROM online phase in Algorithm~\ref{alg:APG-ROM_Online}} & 1    &   $2r^4 + 2r^3 + 2r^2 + 2r$    \\
			& 2    	& $2r$   						\\\cline{2-3} 
			& Total: $flops_{\text{eAPG,Online}}=$	& $2r^4 + 2r^3 + 2r^2 + 4r$		\\\hline
	\end{tabular}
	\bigskip
\end{table}
\begin{table}[t!]
	\centering 
	\caption{Flops required for the online evaluation of the general APG-ROM~\eqref{eq:APG_ROM_general} for a single time step using an explicit Euler integration and Algorithm~\ref{alg:General-APG-ROM_Online}.}
	\label{tab:Flops_General-APG-ROM}
	\begin{tabular}{ll}\hline
		Step in Algorithm~\ref{alg:General-APG-ROM_Online} & Flop Count \\ \hline
		1           & $2rN$        \\
		2           & $(\omega_1 + \omega_2)N$        \\
		3           & $2dN + N$ \\
		4           & $4rN + N$           \\
		5           & $(\omega_1 + \omega_2)N$           \\
		6           & $4dN + N$           \\
		7           & $4rN - 2r + 2r^2$           \\ 
		8           & $2r$           \\ \hline
		Total: $flops_{\text{APG}}=$       & $(6d + 10r + 2\omega_1 + 2\omega_2 + 2)N + 2r^2 - r$ \\\hline
	\end{tabular}
\end{table}

Our investigation reveals that the computational costs associated with the one-time calculation of the G-ROM and the eAPG-ROM coefficients in the offline phase are of order $\mathcal{O}(r^3N)$ and $\mathcal{O}(r^4N)$, respectively. Consequently, the computational effort for the eAPG-ROM is approximately $r$ times higher than that of the G-ROM. In the online phase, the computational complexities for G-ROM and eAPG-ROM are $\mathcal{O}(r^3)$ and $\mathcal{O}(r^4)$, respectively. The cost of the online evaluation of the general APG-ROM for incompressible Navier-Stokes equations is of order $O(rN)$. It is evident that the computational cost of both G-ROM and eAPG-ROM evaluations is significantly lower. However, the computational costs of the offline phase are larger by a factor of $r^2$ or $r^3$ for G-ROM and eAPG-ROM, respectively, than for the evaluation of a single time step of the APG-ROM. We will state/present numerical values for the example in Section~\ref{sec:costs_Karman}.

\section{Numerical example}\label{sec:numerical_example}
We demonstrate the effectiveness of the proposed method through the examination of a three-dimensional simulation of incompressible, single-phase flow past a circular cylinder. 
This benchmark problem is solved using the unsteady Reynolds-Averaged Navier-Stokes (URANS) equations with the $k$--$\omega$ Shear Stress Transport (SST) turbulence model~\cite{Menter1994}. 
The turbulent and periodic time scales of the cylinder wake are well separated, which justifies a statistical treatment of the turbulent fluctuations~\cite{Casimir2019,Casimir2020}. 
The convective flux is discretised by the second-order linear upwind scheme, and the implicit second-order backward-differencing time-marching scheme is applied.
The URANS equations are solved using the OpenFOAM implementation~\cite{Weller1998}. We apply the \texttt{pimpleFoam} solver, which combines the SIMPLE~\cite{Patankar1972} and PISO~\cite{Issa1986} algorithms and is customised for unsteady flow problems.

We present the numerical results, evaluate error estimations from Section~\ref{sec:Error_Eval} and compare the computational costs of the FOM, the G-ROM, and the APG-ROMs in terms of flops and real computational time.
We use the scalar memory term with an arbitrarily selected weight ($w = 1$), its optimized version, and the optimized modified matrix memory term introduced in Section~\ref{sec:MemoryTerm_matrix}. The significance of selecting appropriate weights becomes evident from the accuracy of the obtained results.
\subsection{Flow over a cylinder}\label{sec:Karman}
\begin{figure}[t!]
	\centering
	\def\svgwidth{\columnwidth}
	\includegraphics{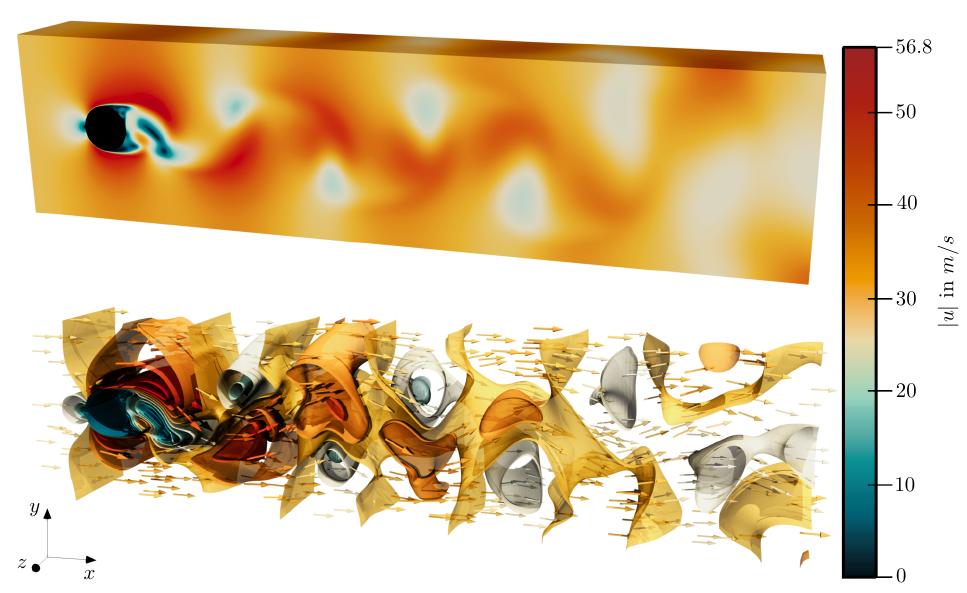}
	\captionsetup{width=\linewidth}
	\caption{Flow over a cylinder: Three-dimensional instantaneous flow field magnitude as a surface plot (top), and a contour representation with gradient glyphs (bottom).} 
	\label{fig:3D_Karman}
\end{figure}
The computational domain is a cuboid measuring $1.7\,\mathrm{m} \times 0.41\,\mathrm{m} \times 0.2\,\mathrm{m}$, containing a cylindrical obstacle of diameter $0.1\,\mathrm{m}$ (see Figure~\ref{fig:3D_Karman}). 
A body-fitted mesh with an average wall-normal resolution of $y^+ \approx 1$ is used, consisting of $1{,}081{,}782$ cells. At the inlet of the computational domain, a Dirichlet condition for velocity is set based on a Reynolds number of $3000$, with a Neumann boundary condition for static pressure. 
At the outlet, a Neumann condition is set for velocity, while a Dirichlet condition with a fixed static pressure of $0\,\mathrm{Pa}$ is used. 
The specified Neumann conditions at both inlet and outlet apply a zero-gradient condition from the internal flow field onto the boundary cell faces.
A no-slip condition is applied at the cylinder walls. For all remaining boundaries, a symmetry condition is applied.

The flow regime is characterised by a von Karman vortex street, which means unsteady periodic vortex shedding at regular intervals.
To accurately capture the temporal evolution of the flow, a time step of $\Delta t = 2.5 \times 10^{-6}\,\mathrm{s}$ is used, corresponding to a maximum CFL number of 0.1. 
Through time step variation, it has been confirmed that the solution is time step independent.
Similarly, the fine near-wall resolution ($y^+ \approx 1$) confirms that the solution is grid independent.

Simulation results are recorded every $1.6 \times 10^{-4}\,\mathrm{s}$ of physical time for the model order reduction. 
The lift coefficient shows periodic oscillations with a frequency of $74.3\,\mathrm{Hz}$, which is half that of the drag coefficient. 
This corresponds to a Strouhal number of 0.247. One complete vortex shedding period is captured using $M = 84$ snapshots for a simulation time  $T_p = 0.01346\,\mathrm{s}$.

\subsubsection{Flow over a cylinder: ROM construction}\label{sec:ROM_Karman}
\begin{figure}[t!]
	\centering
	\def\svgwidth{\columnwidth}
	\includegraphics{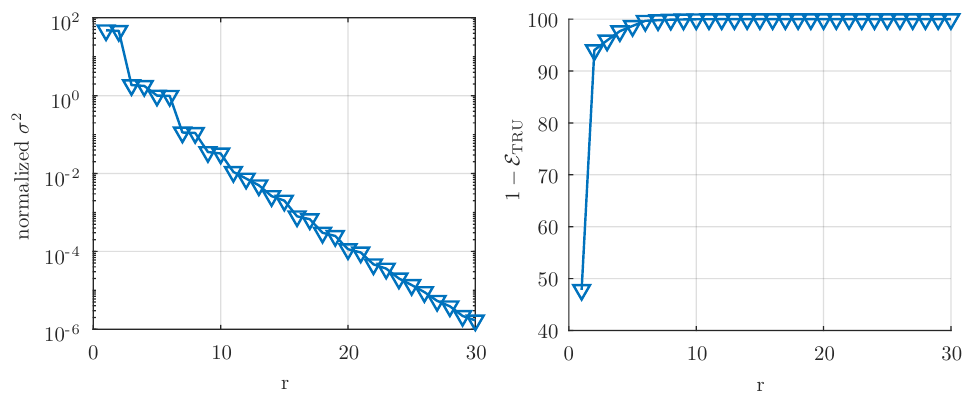}
	\captionsetup{width=\linewidth}
	\caption{Flow over a cylinder: Normalized singular value (left, logarithmic scale) and $1-\mathcal{E}_{\text{TRU}}(r)$ (right) for the first $r = 1,\hdots,30$ trial basis vectors.} 
	\label{fig:Lambdas_Karman}
\end{figure}

The FOM results are interpolated on a uniform Cartesian 3D grid in a post-processing step to simplify the model reduction computation. This grid is defined by cell dimensions $dx = 0.0085\,m$, $dy = 0.050\,m$, and $dz = 0.003125\,m$, leading to a total number of $N_{\text{grid}} = 1084395$ cuboid cells and subsequently, $N = dN_{\text{grid}} = 3253185$ computational nodes. We perform a POD as detailed in Section~\ref{sec:POD}, aiming to extract the dominant flow structures and reduce the computational complexity. We use $r = 8$ trial basis vectors and achieve a truncation error of $\mathcal{E}_{\text{TRU}}(r)=0.099\%$. The singular values are shown in Figure~\ref{fig:Lambdas_Karman} for $r=1,\hdots,30$ and the associated truncation errors~\eqref{sec:TruncationError} for $r=1,\hdots,10$ are shown in Table~\ref{tab:truncation_errors_Karman}. 
\begin{table}[h!]
	\centering
	\caption{Flow over a cylinder: Truncation errors in percent for various $r$.}
	\begin{tabular}{l|llllllllll}
		$r$             & \multicolumn{1}{c}{1}	&	\multicolumn{1}{c}{2}	&	\multicolumn{1}{c}{3}	&	\multicolumn{1}{c}{4}	&	\multicolumn{1}{c}{5}	&	\multicolumn{1}{c}{6}	&	\multicolumn{1}{c}{7}	&	\multicolumn{1}{c}{8}	&	\multicolumn{1}{c}{9}	&	\multicolumn{1}{c}{10} \\ \hline
		$\mathcal{E}_{\text{TRU}}$ in $\%$ & 52.167	&	5.971	&	4.105	&	2.322	&	1.313	&	0.322	&	0.209	&	0.099	&	0.063	&	0.030	 \\
	\end{tabular}\label{tab:truncation_errors_Karman}
\end{table}
\clearpage

\begin{figure}[t!]
	\centering
	\def\svgwidth{\columnwidth}
	\includegraphics{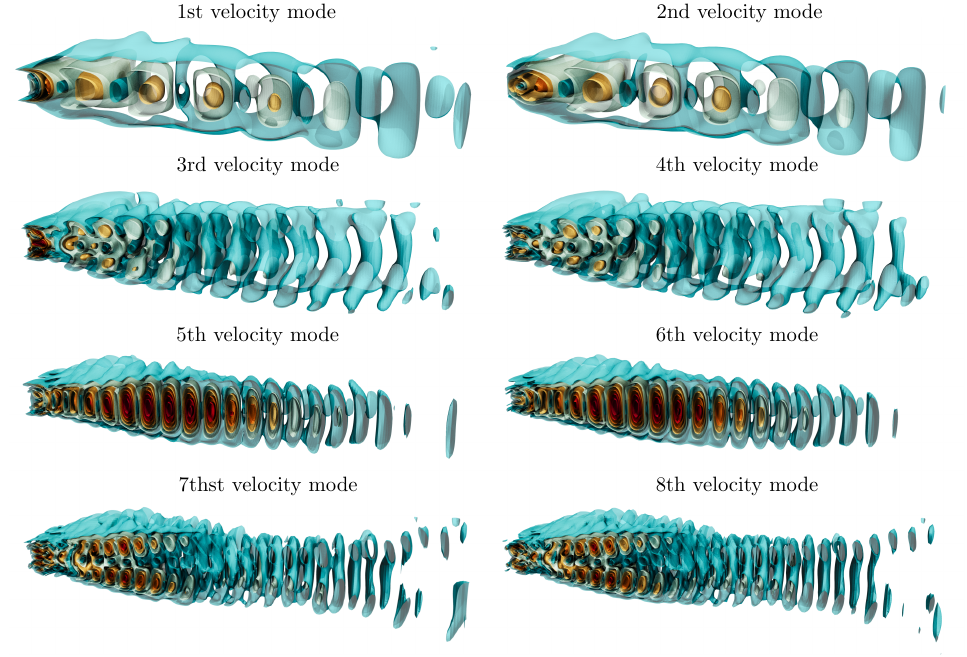}
	\captionsetup{width=\linewidth}
	\caption{Flow over a cylinder: 3D contour plots of the velocity modes $\bm{\phi}_i$, $i=1,\hdots, 8$.} 
	\label{fig:modes_Karman}
\end{figure}

The corresponding first eight 3D trial basis vectors are illustrated in Figure~\ref{fig:modes_Karman}. With these trial basis vectors, we construct the G-ROM and three eAPG-ROMs, specifically the\\
\begin{itemize}
	\item eAPG-ROM with a scalar memory length $\tau$ and an arbitrarily selected weight, where $w=1$;
	\item eAPG-ROM with an optimized scalar memory length $\tau^{\text{opt}}$ and an optimized weight;
	\item eAPG-ROM with a matrix memory length $\bm{T}^{\text{opt}}$ and optimized weights.\\
\end{itemize}
As discussed in Section~\ref{sec:MemoryTerm_matrix} we extend the optimization to include two flow periods.
We do not compute the results for the general APG-ROM, under the assumption that the outputs of the eAPG-ROM and the general APG-ROM are expected to be nearly identical, differing only by negligible numerical errors. We arrive at this conclusion because the eAPG-ROM is rooted in the APG-ROM, which implies that the computational results of the eAPG-ROM are equivalent to those of the general APG-ROM. Given that the computational effort and time required for the general APG-ROM are significantly higher, as demonstrated in Section~\ref{sec:FLOPS}, we deem the additional computations unnecessary.

\subsubsection{Flow over a cylinder: Numerical results}\label{sec:results_Karman}
\begin{figure}[t!]
	\centering
	\def\svgwidth{\columnwidth}
	\includegraphics{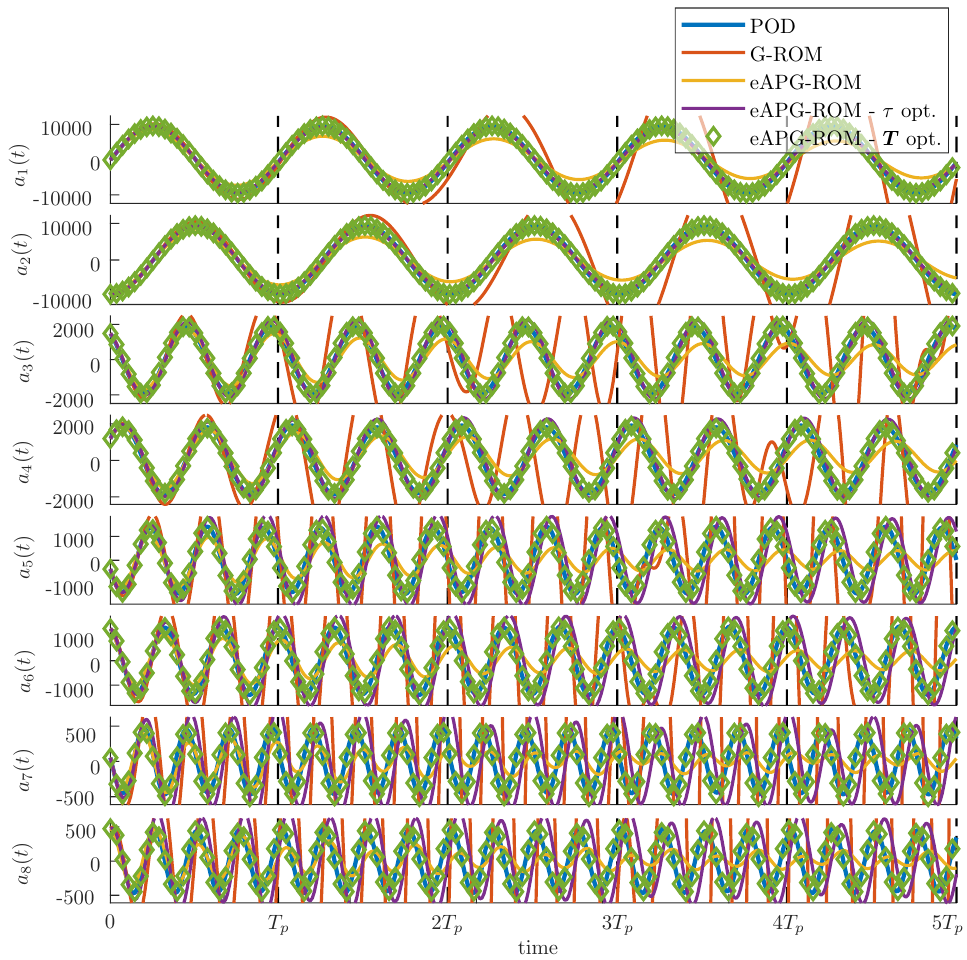}
	\captionsetup{width=\linewidth}
	\caption{Flow over a cylinder: First $8$ modal coefficients $\bm{a}^{\text{POD}}$ (blue), $\bm{a}^{\text{G}}$ (red), $\bm{a}^{\text{eAPG},\tau}$ (yellow), $\bm{a}^{\text{eAPG},\tau,\text{opt}}$ (purple) and $\bm{a}^{\text{eAPG},\bm{T},\text{opt}}$ (green, diamonds) for five periods $T_p$ represented by black dashed lines.} 
	\label{fig:Modal_coeffs_Karman}
\end{figure}

\begin{figure}[t!]
	\centering
	\def\svgwidth{\columnwidth}
	\includegraphics{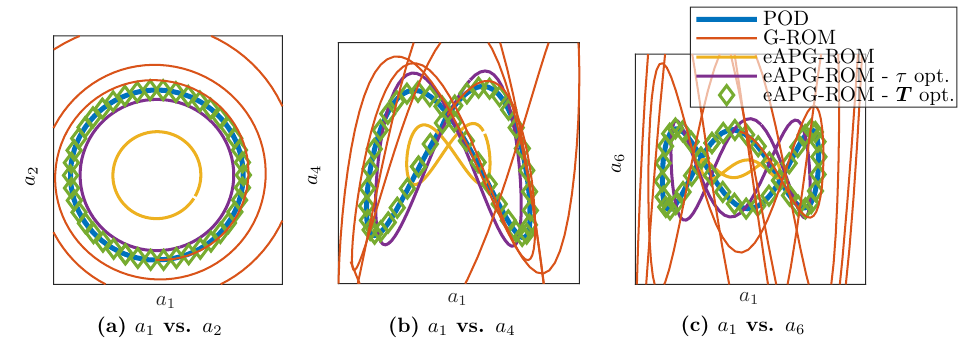}
	\captionsetup{width=\linewidth}
	\caption{Flow over a cylinder: Orbits showing the limit cycle: $\bm{a}^{\text{POD}}$ (blue), $\bm{a}^{\text{G}}$ (red) for the first three periods, whereas $\bm{a}^{\text{eAPG}, \tau}$ (yellow), $\bm{a}^{\text{eAPG}, \tau, \text{opt}}$ (purple) and $\bm{a}^{\text{eAPG}, \bm{T}, \text{opt}}$ (green, diamonds) for the 1000th period. While the standard G-ROM becomes unstable, all eAPG-ROMs converge to a stable limit cycle.} 
	\label{fig:Orbits_Karman}
\end{figure}

We present a discussion of the numerical results obtained from solving the ROMs using numerical integration methods. We use the Dormand-Prince integration scheme (implemented as \textit{ode45} in \textsc{Matlab} \cite{Dormand1980}) and begin with a comparison of the modal coefficients. Figures~\ref{fig:Modal_coeffs_Karman} and~\ref{fig:Orbits_Karman} illustrate the first eight resulting modal coefficients over five periods $T_p$ and the limit cycles for some selected modal coefficients, respectively. Here, $\bm{a}^{\text{POD}}$ serves as the reference that we aim to replicate. It is evident that the results obtained from the G-ROM $\bm{a}^{\text{G}}$ show a deviation that grows over time, eventually leading to instability. This instability is also evident from the orbits shown in Figure~\ref{fig:Orbits_Karman}, where only the initial three periods are displayed. The eAPG-ROM with scalar memory length weight $w=1$, denoted by $\bm{a}^{\text{eAPG},\tau}$, yields improved results compared to the standard G-ROM. 
Despite observing the system for 1000 periods (see Figure~\ref{fig:Orbits_Karman}), no signs of instability were detected. However, the system fails to replicate the original limit cycle accurately. The eAPG-ROM with optimized scalar memory length $\bm{a}^{\text{eAPG},\tau,\text{opt}}$ shows a comparable trajectory, although with improved results. Notably, the limit cycle has a smaller deviation from the reference in comparison to the eAPG-ROM without optimization. The best results are achieved with the eAPG-ROM and an optimized matrix memory length, denoted by $\mathbf{a}^{\text{eAPG},\mathbf{T}, \text{opt}}$, showing a good agreement with the reference. The limit cycle closely approximates the reference with minimal error even for the 1000th period in this case.

\begin{figure}[b!]
	\centering
	\def\svgwidth{\columnwidth}
	\includegraphics{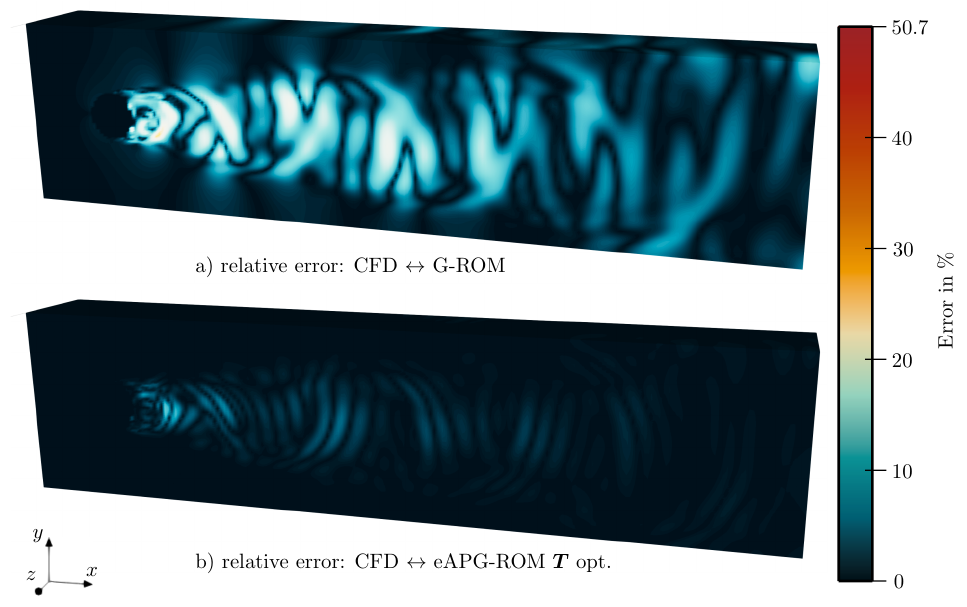}
	\captionsetup{width=\linewidth}
	\caption{Flow over a cylinder: Surface of the 3D flow: Instantaneous (a) relative error of the reconstructed velocity field magnitude $\bm{u}^{\text{G}}$ (G-ROM) with respect to the CFD results and (b) relative error of the reconstructed velocity field magnitude $\bm{u}^{\text{eAPG}, \bm{T}, \text{opt}}$ (eAPG-ROM with optimized matrix memory length) with respect to the CFD results for the $57$th time step. Relative errors are normalized to the reference velocity $u_\text{ref}$. The time step was selected to show the maximal relative error of the eAPG-ROM with optimized matrix memory length.} 
	\label{fig:velocity_Karman}
\end{figure}

Additionally, we show instantaneous relative flow field errors in Figure~\ref{fig:velocity_Karman} for the $57$th time step to provide spatial information of the reconstructed flow fields. 
The time step was selected to show the maximal relative error of the eAPG-ROM in one period with optimized matrix memory length. These flow fields are computed on the uniform Cartesian grid. Figures~\ref{fig:velocity_Karman}a and \ref{fig:velocity_Karman}b show the relative errors between the original flow field and those generated by the G-ROM and the eAPG-ROM with optimized matrix memory length for the same time step. The relative errors are normalized to the reference velocity $u_{\text{ref}} = 30\,\text{m/s}$, which is the inlet velocity. The G-ROM shows local errors of up to approximately $50.7\%$ for this time step. The eAPG-ROM shows significantly lower local errors, with a maximum of $8.4\%$. The error for the G-ROM increases for subsequent time steps due to its instability (see Figure~\ref{fig:Orbits_Karman}), leading to a growing mismatch between the reconstructed and original flow field. In contrast, the local errors for the eAPG-ROM stay below $8.4\%$ for all times. As the eAPG-ROM converges to a stable limit cycle as shown in Figure~\ref{fig:Orbits_Karman}, it maintains bounded errors even as the simulation continues for long times.

\begin{table}[t!]
	\centering
	\caption{Flow over a cylinder: ROM, Total and Reconstruction errors in percent for the investigated ROMs.}
	\begin{tabular}{l|ccc}
		ROMs             & $\mathcal{E}_{\text{ROM}}$ in $\%$ &	$\mathcal{E}_{\text{Total}}$ in $\%$ & $\mathcal{E}_{\text{REC}}$ in $\%$ \\ \hline
		G-ROM & $7.8627$	&	$7.9616$	&	$3.1186\cdot 10^{-4}$		 \\
		eAPG-ROM & $3.6041$	&	$3.7029$	&	$2.1269\cdot 10^{-4}$		 \\
		eAPG-ROM - $\tau$ opt. & $0.5515$	&	$0.6503$	&	$8.9135\cdot 10^{-5}$		 \\
		eAPG-ROM - $\bm{T}$ opt. & $0.0093$	&	$0.1081$	&	$3.6349\cdot 10^{-5}$		 \\
	\end{tabular}\label{tab:ROM_errors_Karman}
\end{table}
\begin{figure}[t!]
	\centering
	\def\svgwidth{\columnwidth}
	\includegraphics{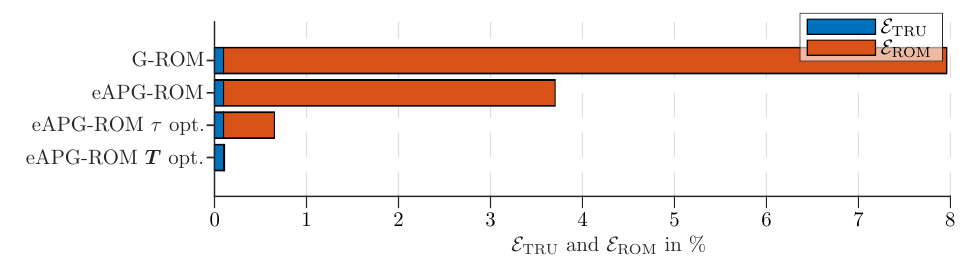}
	\captionsetup{width=\linewidth}
	\caption{Flow over a cylinder: Truncation and ROM Errors $\mathcal{E}_{\text{TRU}}$ and $\mathcal{E}_{\text{ROM}}$ for the investigated ROMs in percent.} 
	\label{fig:modalErrors_Karman}
\end{figure}
\begin{figure}[t!]
	\centering
	\def\svgwidth{\columnwidth}
	\includegraphics{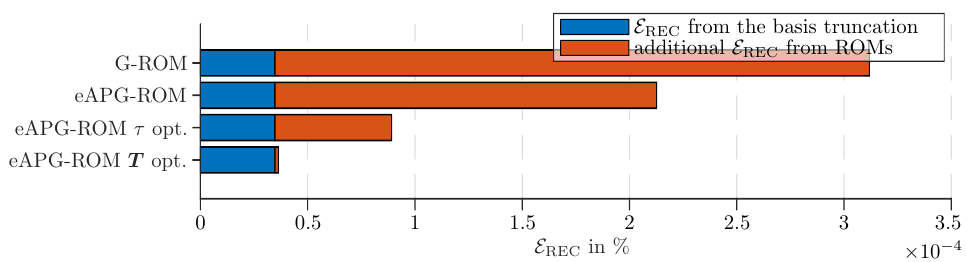}
	\captionsetup{width=\linewidth}
	\caption{Flow over a cylinder: Reconstruction Error $\mathcal{E}_{\text{REC}}$ for the trial basis truncation and the investigated ROMs in percent.} 
	\label{fig:recErrors_Karman}
\end{figure}
The scalar error metrics associated with the ROMs are crucial for a quick comparison of their performances. We calculate the errors introduced in Section~\ref{sec:Error_Eval} and show them in Figures \ref{fig:modalErrors_Karman} and \ref{fig:recErrors_Karman} and in Table~\ref{tab:ROM_errors_Karman}. The analysis clearly indicates that the errors associated with the ROMs, including the ROM error, total error, and reconstruction error, are significantly higher for the standard G-ROM. Specifically, the ROM error $\mathcal{E}_{\text{ROM}}$ associated with the G-ROM is observed to be nearly two orders of magnitude higher than the error resulting from basis truncation $\mathcal{E}_{\text{TRU}}$ alone. Even by incorporating an arbitrarily selected weight in the eAPG-ROM, there is already a noticeable improvement in the accuracy of the results.

Further refinement is achieved through the optimization of the eAPG-ROM, leading to a ROM error that is approximately five times bigger than the truncation error. The most substantial improvement is observed with the eAPG-ROM that uses an optimized matrix memory length, where the resulting error is an order of magnitude lower than the basis truncation error. This trend is also reflected in the reconstruction errors $\mathcal{E}_{\text{REC}}$, where the largest errors are observed in the G-ROM, while the smallest errors are associated with the eAPG-ROM with optimized matrix memory length. Specifically, in the latter case, the additional reconstruction error is approximately an order of magnitude lower than the reconstruction error introduced by the basis truncation $\mathcal{E}_{\text{REC}} = 3.4745\cdot 10^{-5}\%$ (compare Figure~\ref{fig:recErrors_Karman}). This observation is consistent with the error analysis of the temporal behavior presented in Figures~\ref{fig:Modal_coeffs_Karman} and~\ref{fig:Orbits_Karman} and the instantaneous flow fields in Figure~\ref{fig:velocity_Karman}.

\subsubsection{Flow over a cylinder: Computational Cost}\label{sec:costs_Karman}
\begin{figure}[t!]
	\centering
	\def\svgwidth{\columnwidth}
	\includegraphics{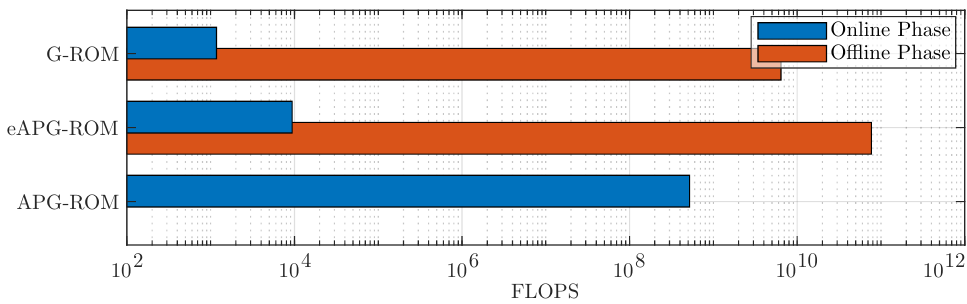}
	\captionsetup{width=\linewidth}
	\caption{Flow over a cylinder: Flops for the G-ROM, eAPG-ROM and APG-ROM in the online and offline phases with a logarithmic scale.} 
	\label{fig:FLOPS_Karman}
\end{figure}

Finally, we investigate the numerical costs of constructing and solving the ROMs. First, we analyze the theoretical number of flops for the Algorithms~\ref{alg:G-ROM_Offline},~\ref{alg:G-ROM_Online},~\ref{alg:APG-ROM_Offline},~\ref{alg:APG-ROM_Online}, and~\ref{alg:General-APG-ROM_Online}. Crucial to our computational approach is the use of second-order central derivatives wherever feasible, supplemented by second-order forward or backward derivatives at the boundaries of the computational domain. Specifically, we assume the computational cost for computing the first and second spatial derivatives to be $\omega_1 N$ and $\omega_2 N$, respectively, where $\omega_1 = 12$ and $\omega_2 = 18$. This stems from the interpolation performed onto the structured Cartesian uniform grid, where the computation of the first derivative requires two function evaluations, while the computation of the second derivative requires three, along with the corresponding numerical operations. Flops are summarized in Figure~\ref{fig:FLOPS_Karman} and Table~\ref{tab:FLOPS_Karman}.
\begin{table}[t!]
	\centering
	\caption{Flow over a cylinder: Flops for the G-ROM, eAPG-ROM and APG-ROM in the online and offline phases.}
		\begin{tabular}{l|l|l}
			ROMs             		& offline phase 	& online phase \\ \hline
			G-ROM 					& \textit{Alg.~\ref{alg:G-ROM_Offline}:} $6,402,267,496$		& \textit{Alg.~\ref{alg:G-ROM_Online}:}	$1,176$ \\
			eAPG-ROM 				& \textit{Alg.~\ref{alg:APG-ROM_Offline}:} $76,745,882,071$	& \textit{Alg.~\ref{alg:APG-ROM_Online}:} $9,376$	\\
			APG-ROM 				& -															& \textit{Alg.~\ref{alg:General-APG-ROM_Online}:} $520,509,720$ \\
		\end{tabular}\label{tab:FLOPS_Karman}
\end{table}

The analysis reveals notable differences in computational costs for both the offline and online phases. Specifically, the computational cost for the offline phase of the eAPG-ROM is observed to be approximately $11.99$ times higher than that of the G-ROM. In the online phase, the eAPG-ROM requires $~7.97$ times the computational cost of the G-ROM. Note that, as the dimension $r$ of the ROMs increases, the computational cost in the online phase converges towards a factor of $r$. This is due to the computational complexities of the G-ROM and eAPG-ROM as discussed in Section~\ref{sec:FLOPS}, which are $\mathcal{O}(r^3)$ and $\mathcal{O}(r^4)$, respectively. Moreover, despite the marginally increased computational costs associated with the eAPG-ROM, significant enhancements in accuracy and stability, as shown in the previous section, are evident. Furthermore, a notable growth is visible when comparing the computational costs for a single evaluation of the online phases of the general APG-ROM and the eAPG-ROM. Specifically, the general APG-ROM demands approximately $55,515$ times more flops per evaluation.

In evaluating the computational efficiency of the ROMs under consideration, it is necessary to analyze both theoretical computational costs and real computational time. We measured the simulation time to solve the FOM and execute the online phases of the G-ROM and eAPG-ROM for one period (see Table~\ref{tab:Simulation_Time_Karman}). 
\begin{table}[h!]
	\centering
	\caption{Flow over a cylinder: Measured simulation time of the FOM, G-ROM and eAPG-ROM with optimized matrix memory length for one flow period $T_p$. The G-ROM and eAPG-ROM use the Dormand-Prince integration scheme.}
	\label{tab:Simulation_Time_Karman}
	\begin{threeparttable}
		\begin{tabular}{l|l}
			Simulation             								& Simulation time in s 	 \\ \hline
			FOM\tnote{1} 										& $20152.5$ \\
			G-ROM: online phase\tnote{2} 						& $2.7687\cdot 10^{-7}$	\\
			eAPG-ROM - $\bm{T}$ opt.: online phase\tnote{2} 	& $2.1185\cdot 10^{-6}$\\
		\end{tabular}
		\begin{tablenotes}
			\item[1] Solved using OpenFOAM, 2 $\times$ Intel Xeon Gold 5120 CPU @ 2.20GHz (56 threads)
			\item[2] Solved using \textsc{Matlab}, Intel Core i5-1250p @ 4.40 GHz (16 threads)
		\end{tablenotes}
	\end{threeparttable}
\end{table}

Due to inherent similarities, the computation of results and corresponding time measurements for the general APG-ROM have been omitted as explained in Section~\ref{sec:ROM_Karman}.
Moreover, the theoretical computational costs are considered to be sufficient to conclude that the real computational times will indeed surpass those of the eAPG-ROM by magnitudes comparable to the theoretical computational costs themselves.

The measured simulation times show a speedup of the G-ROM and eAPG-ROM compared to the FOM of $9.512\cdot 10^{9}$ and $7.278\cdot 10^{10}$, respectively. Note that the FOM was executed on a high-performance computing server, whereas the ROMs were solved on a standard consumer-grade personal computer. The observed speedup factor for the G-ROM compared to the eAPG-ROM is $7.6519$. Although this factor is slightly smaller than the theoretical speedup based on flops, the deviation with the expected theoretical performance is minimal.

\section{Conclusion and Outlook}\label{sec:conclusion_and_outlook}
This paper has presented a novel contribution in the form of an efficient implementation of the Adjoint Petrov-Galerkin reduced order model (APG-ROM) tailored specifically for fluid flows described by the incompressible Navier-Stokes equations. The proposed efficient APG-ROM (eAPG-ROM) builds on the foundations laid by the general APG-ROM introduced in the work by Parish et al.~\cite{Parish2020}. Rooted in the Mori-Zwanzig Formalism, the general APG-ROM introduces a novel method to compute trial basis vectors for the Petrov-Galerkin projection. However, a notable drawback is the temporal variation inherent in these trial basis vectors, that poses a challenge for systems with a high number of degrees of freedom. A crucial consequence of this temporal variation is the necessity for recalculating the trial basis vectors at each time step during the numerical integration of the APG-ROM. While the general APG-ROM exhibits promising results in terms of accuracy and stability, it requires computations in the original spatial domain of the FOM, which compromises the reduction efforts. 

The proposed eAPG-ROM exploits the polynomial nature of the incompressible Navier-Stokes equations. The computational strategy involves a separation of tasks into distinct offline and online phases. The offline phase, conducted in the full spatial domain, contains the computation of the ROMs coefficients with the trial basis vectors. However, the subsequent online evaluation of the eAPG-ROM occurs in the dimension of the number of truncated coarse-scale trial basis vectors $r$. This strategic partitioning results in a substantial reduction in computational costs for the online evaluation compared to the evaluation of the general APG-ROM. Furthermore, we introduced a novel, straightforward, yet efficient approach designed to augment the number of degrees of freedom in evaluating the memory length - an important factor influencing the stability and accuracy of the APG-ROM. Our demonstration employed a data-driven method to determine this augmented memory length, with the objective of optimizing the stability and accuracy of the eAPG-ROM. Numerical experiments, conducted on the flow around a cylinder, demonstrate the superiority of the eAPG-ROM and significantly improved results compared to a standard Galerkin ROM (G-ROM). When examining the computational costs of the online phase of the eAPG-ROM, we can show that the associated costs are approximately $r$ times higher compared to those caused by the G-ROM. This observation is derived from our evaluation of counted flops. Despite this increase in computational effort, we stress that the overall costs remain significantly lower when compared to the computational demands imposed by the general APG-ROM when applied to the incompressible Navier-Stokes equations. This factor may undergo variations with higher-order integration schemes. However, even under such circumstances, the computational expenses are constrained in the range of $\mathcal{O}(r)$. This shows the practicality and efficiency of the proposed approach, making it a viable option for real-world applications. In our future research, the focus will be directed towards the application of this technique to more complex systems, e.g. real 3D centrifugal pumps [2], and parametric APG-ROMs for real-time control problems.

\section*{Acknowledgement}
Funded by the Deutsche Forschungsgemeinschaft (DFG, German Research Foundation) – Project-ID 422037413 – TRR 287.

% \section*{References}
\bibliographystyle{ieeetr}
\bibliography{bibtex}

\begin{thebibliography}{10}

\bibitem{John2010}
T.~John, M.~Guay, N.~Hariharan, and S.~Naranayan, ``{POD-based observer for
  estimation in Navier-Stokes flow},'' {\em {Computers and Chemical
  Engineering}}, vol.~34, no.~6, pp.~965 -- 975, 2010.

\bibitem{Seoane2020}
M.~Seoane, P.~D. Ledger, A.~J. Gil, S.~Zlotnik, and M.~Mallett, ``{A combined
  reduced order-full order methodology for the solution of 3D
  magneto-mechanical problems with application to magnetic resonance imaging
  scanners},'' {\em International Journal for Numerical Methods in
  Engineering}, vol.~121, no.~16, pp.~3529--3559, 2020.

\bibitem{Bergmann2008}
M.~Bergmann and L.~Cordier, ``{Optimal control of the cylinder wake in the
  laminar regime by trust-region methods and POD reduced-order models},'' {\em
  Journal of Computational Physics}, vol.~8, pp.~7813--7840, 2008.

\bibitem{Deane1991}
A.~Deane, I.~G. Kevrekidis, G.~Karniadakis, and S.~A. Orszag,
  ``{Low‐dimensional models for complex geometry flows: Application to
  grooved channels and circular cylinders},'' {\em Physics of Fluids}, vol.~3,
  pp.~2337--2354, 1991.

\bibitem{Liberge2010}
E.~Liberge and A.~Hamdouni, ``{Reduced order modelling method via proper
  orthogonal decomposition (POD) for flow around an oscillating cylinder},''
  {\em Journal of Fluids and Structures}, vol.~26, no.~2, pp.~292 -- 311, 2010.

\bibitem{Gunder2018}
T.~Gunder, A.~Sehlinger, R.~Skoda, and M.~M\"onnigmann, ``{Sensor placement for
  reduced-order model-based observers in hydraulic fluid machinery},'' {\em
  IFAC-PapersOnLine}, vol.~51, no.~13, pp.~414--419, 2018.

\bibitem{Sommer2023}
K.~D. Sommer, L.~Reineking, Y.~P. Ravichandran, R.~Skoda, and M.~Mönnigmann,
  ``Estimating flow fields with reduced order models,'' {\em Heliyon}, vol.~9,
  no.~11, p.~e20930, 2023.

\bibitem{Berner2017}
M.~O. Berner, F.~Sudbrock, V.~Scherer, and M.~M\"onnigmann, ``{POD and
  Galerkin-based reduction of a wood chip drying model},'' {\em
  IFAC-PapersOnLine}, vol.~50, no.~1, pp.~6619--6623, 2017.

\bibitem{Berner2020}
M.~O. Berner, V.~Scherer, and M.~M\"onnigmann, ``An observer for partially
  obstructed wood particles in industrial drying processes,'' {\em Computers \&
  Chemical Engineering}, vol.~141, p.~107013, 2020.

\bibitem{Meyer2017}
D.~S. Meyer, B.~T. Helenbrook, and M.-C. Cheng, ``{Proper Orthogonal
  Decomposition-based reduced basis element thermal modeling of integrated
  circuits},'' {\em International Journal for Numerical Methods in
  Engineering}, vol.~112, no.~5, pp.~479--500, 2017.

\bibitem{Akhtar2012}
I.~Akhtar, Z.~Wang, J.~Borggaard, and T.~Iliescu, ``A new closure strategy for
  proper orthogonal decomposition reduced-order models,'' {\em Journal of
  Computational and Nonlinear Dynamics}, vol.~7, no.~3, p.~034503, 2012.

\bibitem{Baiges2015}
J.~Baiges, R.~Codina, and S.~Idelsohn, ``Reduced-order subscales for {POD}
  models,'' {\em Computer Methods in Applied Mechanics and Engineering},
  vol.~291, pp.~173--196, 2015.

\bibitem{Ahmed2021}
S.~E. Ahmed, S.~Pawar, O.~San, A.~Rasheed, T.~Iliescu, and B.~R. Noack, ``On
  closures for reduced order models—a spectrum of first-principle to
  machine-learned avenues,'' {\em Physics of Fluids}, vol.~33, no.~9,
  p.~091301, 2021.

\bibitem{Couplet2003}
M.~Couplet, P.~Sagaut, and C.~Basdevant, ``{Intermodal energy transfers in a
  proper orthogonal decomposition–Galerkin representation of a turbulent
  separated flow},'' {\em Journal of Fluid Mechanics}, vol.~491, p.~275–284,
  2003.

\bibitem{Bergmann2009}
M.~Bergmann, C.-H. Bruneau, and A.~Iollo, ``{Enablers for robust POD models},''
  {\em Journal of Computational Physics}, vol.~228, no.~2, pp.~516--538, 2009.

\bibitem{Sagaut2006}
P.~Sagaut, {\em {Large Eddy Simulation for Incompressible Flows}}.
\newblock Berlin: Scientific Computation, Springer-Verlag, 3~ed., 2006.

\bibitem{Aubry1988}
N.~Aubry, P.~Holmes, J.~L. Lumley, and E.~Stone, ``The dynamics of coherent
  structures in the wall region of a turbulent boundary layer,'' {\em Journal
  of Fluid Mechanics}, vol.~192, p.~115–173, 1988.

\bibitem{Ullmann2010}
S.~Ullmann and J.~Lang, ``{A POD-Galerkin reduced model with updated
  coefficients for Smagorinsky LES},'' in {\em V European conference on
  computational fluid dynamics, ECCOMAS CFD}, p.~2010, 2010.

\bibitem{Xie2016}
X.~Xie, D.~Wells, Z.~Wang, and T.~Iliescu, ``Approximate deconvolution reduced
  order modeling,'' {\em Computer Methods in Applied Mechanics and
  Engineering}, vol.~313, pp.~512--534, 2017.

\bibitem{Xie2018}
X.~Xie, M.~Mohebujjaman, L.~G. Rebholz, and T.~Iliescu, ``Data-driven filtered
  reduced order modeling of fluid flows,'' {\em SIAM Journal on Scientific
  Computing}, vol.~40, no.~3, pp.~B834--B857, 2018.

\bibitem{Wells2017}
D.~Wells, Z.~Wang, X.~Xie, and T.~Iliescu, ``An evolve-then-filter regularized
  reduced order model for convection-dominated flows,'' {\em International
  Journal for Numerical Methods in Fluids}, vol.~84, no.~10, pp.~598--615,
  2017.

\bibitem{Sabetghadam2012}
F.~Sabetghadam and A.~Jafarpour, ``{$\textalpha$ Regularization of the
  POD-Galerkin dynamical systems of the Kuramoto–Sivashinsky equation},''
  {\em Applied Mathematics and Computation}, vol.~218, no.~10, pp.~6012--6026,
  2012.

\bibitem{Themistoklis2009}
T.~P. Sapsis and P.~F. Lermusiaux, ``Dynamically orthogonal field equations for
  continuous stochastic dynamical systems,'' {\em Physica D: Nonlinear
  Phenomena}, vol.~238, no.~23, pp.~2347--2360, 2009.

\bibitem{Resseguier2015}
V.~Resseguier, E.~M{\'e}min, and B.~Chapron, ``Stochastic fluid dynamic model
  and dimensional reduction,'' in {\em Ninth International Symposium on
  Turbulence and Shear Flow Phenomena}, pp.~649--654, 2015.

\bibitem{Couplet2005}
M.~Couplet, C.~Basdevant, and P.~Sagaut, ``{Calibrated reduced-order
  POD-Galerkin system for fluid flow modelling},'' {\em {Journal of
  Computational Physics}}, vol.~207, pp.~192--220, 2005.

\bibitem{Buffoni2006}
M.~Buffoni, S.~Camarri, A.~Iollo, and M.~V. Salvetti, ``Low-dimensional
  modelling of a confined three-dimensional wake flow,'' {\em Journal of Fluid
  Mechanics}, vol.~569, p.~141–150, 2006.

\bibitem{Perret2006}
E.~C. Laurent~Perret and J.~Delville, ``{Polynomial identification of POD based
  low-order dynamical system},'' {\em Journal of Turbulence}, vol.~7, p.~N17,
  2006.

\bibitem{Zucatti2021}
V.~Zucatti and W.~Wolf, ``Data-driven closure of projection-based reduced order
  models for unsteady compressible flows,'' {\em Computer Methods in Applied
  Mechanics and Engineering}, vol.~386, p.~114120, 2021.

\bibitem{San2015}
O.~San and T.~Iliescu, ``A stabilized proper orthogonal decomposition
  reduced-order model for large scale quasigeostrophic ocean circulation,''
  {\em Advances in Computational Mathematics}, vol.~41, pp.~1289--1319, 2015.

\bibitem{Pawar2019}
S.~Pawar, S.~M. Rahman, H.~Vaddireddy, O.~San, A.~Rasheed, and P.~Vedula, ``{A
  deep learning enabler for nonintrusive reduced order modeling of fluid
  flows},'' {\em Physics of Fluids}, vol.~31, no.~8, p.~085101, 2019.

\bibitem{Mjalled2023a}
R.~Jendersie, A.~Mjalled, X.~Lu, L.~Reineking, A.~Kharaghani, M.~M\"onnigmann,
  and C.~Lessig, ``{NeuroPNM: Model Reduction of Pore Network Models using
  Neural Networks},'' {\em Particuology}, vol.~86, pp.~239--251, 2023.

\bibitem{Mjalled2023b}
A.~Mjalled, E.~Torres, and M.~Mönnigmann, ``Reduced-order modeling framework
  using two-level neural networks,'' {\em PAMM}, vol.~23, no.~2, p.~e202300061,
  2023.

\bibitem{Borggaard2008}
J.~Borggaard, A.~Duggleby, A.~Hay, T.~Iliescu, and Z.~Wang, ``Reduced-order
  modeling of turbulent flows,'' in {\em Proceedings of MTNS}, vol.~2008, 2008.

\bibitem{Wang2012b}
Z.~Wang, I.~Akhtar, J.~Borggaard, and T.~Iliescu, ``Proper orthogonal
  decomposition closure models for turbulent flows: A numerical comparison,''
  {\em Computer Methods in Applied Mechanics and Engineering}, vol.~237-240,
  pp.~10--26, 2012.

\bibitem{Stabile2019}
G.~Stabile, F.~Ballarin, G.~Zuccarino, and G.~Rozza, ``A reduced order
  variational multiscale approach for turbulent flows,'' {\em Advances in
  Computational Mathematics}, vol.~45, pp.~2349--2368, 2019.

\bibitem{Reyes2020}
R.~Reyes and R.~Codina, ``Projection-based reduced order models for flow
  problems: A variational multiscale approach,'' {\em Computer Methods in
  Applied Mechanics and Engineering}, vol.~363, p.~112844, 2020.

\bibitem{Mou2021}
C.~Mou, B.~Koc, O.~San, L.~G. Rebholz, and T.~Iliescu, ``Data-driven
  variational multiscale reduced order models,'' {\em Computer Methods in
  Applied Mechanics and Engineering}, vol.~373, p.~113470, 2021.

\bibitem{Carlberg2011}
K.~Carlberg, C.~Bou-Mosleh, and C.~Farhat, ``{Efficient non-linear model
  reduction via a least-squares Petrov–Galerkin projection and compressive
  tensor approximations},'' {\em International Journal for Numerical Methods in
  Engineering}, vol.~86, no.~2, pp.~155--181, 2011.

\bibitem{Carlberg2017}
K.~Carlberg, M.~Barone, and H.~Antil, ``{Galerkin v. least-squares
  Petrov–Galerkin projection in nonlinear model reduction},'' {\em Journal of
  Computational Physics}, vol.~330, pp.~693--734, 2017.

\bibitem{Kragel2005}
B.~Kragel, {\em {Streamline Diffusion POD Models in Optimization}}.
\newblock PhD thesis, Universit{\"a}t Trier, 2005.

\bibitem{Giere2015}
S.~Giere, T.~Iliescu, V.~John, and D.~Wells, ``Supg reduced order models for
  convection-dominated convection–diffusion–reaction equations,'' {\em
  Computer Methods in Applied Mechanics and Engineering}, vol.~289,
  pp.~454--474, 2015.

\bibitem{Reineking2022}
L.~Reineking, K.~Sommer, Y.~P. Ravichandran, R.~Skoda, and M.~Mönnigmann,
  ``Long-term stable reduced models for hydraulic systems governed by reynolds
  averaged navier-stokes equations,'' {\em IFAC-PapersOnLine}, vol.~55, no.~7,
  pp.~254--259, 2022.

\bibitem{Parish2020}
E.~J. Parish, C.~R. Wentland, and K.~Duraisamy, ``{The Adjoint
  Petrov–Galerkin method for non-linear model reduction},'' {\em Computer
  Methods in Applied Mechanics and Engineering}, vol.~365, p.~112991, 2020.

\bibitem{Hughes2000}
T.~J.~R. Hughes, {\em The Finite Element Method: Linear Static and Dynamic
  Finite Element Analysis}.
\newblock Dover Publications, 2000.

\bibitem{Sirovich1987a}
L.~Sirovich, ``{Turbulence and the dynamics of coherent structures. {P}art {I}:
  Coherent structures},'' {\em Quarterly of applied Mathematics}, vol.~45,
  no.~3, pp.~561--571, 1987.

\bibitem{Zhaojun2000}
Z.~Bai, J.~Demmel, J.~Dongarra, A.~Ruhe, and H.~van~der Vorst, {\em Templates
  for the solutions of algebraic eigenvalue problems a practical guide}.
\newblock Philadelphia: Society for Industrial and Applied Mathematics, 2000.

\bibitem{Chorin2009}
A.~J. Chorin and O.~H. Hald, {\em Stochastic Tools in Mathematics and Science},
  vol.~2.
\newblock Springer, 2009.

\bibitem{Zwanzig1973}
R.~Zwanzig, ``{Nonlinear generalized Langevin equations},'' {\em Journal of
  Statistical Physics}, vol.~9, no.~3, pp.~215--220, 1973.

\bibitem{Mori1965}
H.~Mori, ``{Transport, Collective Motion, and Brownian Motion*)},'' {\em
  Progress of Theoretical Physics}, vol.~33, no.~3, pp.~423--455, 1965.

\bibitem{Parish2017}
E.~J. Parish and K.~Duraisamy, ``{Non-Markovian closure models for large eddy
  simulations using the Mori-Zwanzig formalism},'' {\em Phys. Rev. Fluids},
  vol.~2, p.~014604, 2017.

\bibitem{Golub2013}
G.~H. Golub and C.~F. {van Loan}, {\em Matrix Computations:}, ch.~2.4.
\newblock The Johns Hopkins University Press, fourth~ed., 2013.

\bibitem{Dormand1980}
J.~Dormand and P.~Prince, ``{A family of embedded Runge-Kutta formulae},'' {\em
  Journal of Computational and Applied Mathematics}, vol.~6, no.~1, pp.~19--26,
  1980.

\bibitem{Addison1993}
C.~Addison, J.~Allwright, N.~Binsted, N.~Bishop, B.~Carpenter, P.~Dalloz,
  D.~Gee, V.~Getov, T.~Hey, R.~Hockney, M.~Lemke, J.~Merlin, M.~Pinches,
  C.~Scott, and I.~Wolton, ``{The Genesis distributed-memory benchmarks. Part
  1: Methodology and general relativity benchmark with results for the SUPRENUM
  computer},'' {\em Concurrency: Practice and Experience}, vol.~5, no.~1,
  pp.~1--22, 1993.

\bibitem{Ueberhuber1997}
C.~W. Ueberhuber, {\em Numerical Computation 1}.
\newblock Heidelberg: Springer-Verlag Berlin, 1997.

\bibitem{Menter1994}
F.~R. Menter, ``Two-equation eddy-viscosity turbulence models for engineering
  applications,'' {\em AIAA Journal}, vol.~32, no.~8, pp.~1598--1605, 1994.

\bibitem{Casimir2019}
N.~Casimir, Z.~Xiangyuan, G.~Ludwig, and R.~Skoda, ``{Assessment of statistical
  eddy-viscosity turbulence models for unsteady flow at part and overload
  operation of centrifugal pumps},'' in {\em Proceedings of 13th European
  Conference on Turbomachinery Fluid Dynamics and Thermodynamics (ETC’13),
  Lausanne, Switzerland}, p.~13, 2019.

\bibitem{Casimir2020}
N.~Casimir, X.~Zhu, M.~Hundshagen, G.~Ludwig, and R.~Skoda, ``Numerical study
  of rotor-stator interaction of a centrifugal pump at part load with special
  emphasis on unsteady blade load,'' {\em Journal of Fluids Engineering},
  vol.~142, no.~8, p.~30, 2020.

\bibitem{Weller1998}
H.~G. Weller, G.~Tabor, H.~Jasak, and C.~Fureby, ``A tensorial approach to
  computational continuum mechanics using object-oriented techniques,'' {\em
  Computers in Physics}, vol.~12, no.~6, pp.~620--631, 1998.

\bibitem{Patankar1972}
S.~V. Patankar and D.~B. Spalding, ``{A calculation procedure for heat, mass
  and momentum transfer in three-dimensional parabolic flows},'' {\em
  International Journal of Heat and Mass Transfer}, vol.~15, no.~3,
  pp.~1787--1806, 1972.

\bibitem{Issa1986}
R.~Issa, ``{Solution of the implicitly discretised fluid flow equations by
  operator-splitting},'' {\em Journal of Computational Physics}, vol.~62,
  no.~1, pp.~40--65, 1986.

\end{thebibliography}

\appendix

\section{Coarse-scale projection}\label{ap:Coarse-scale_Projection}
Applying the coarse-scale projection operator $\tilde{\bm{\Pi}}$ to project the FOM~\eqref{eq:FOM} onto the resolved coarse scales yields
\begin{align}
	&\frac{\partial}{\partial t} \tilde{\bm{\Pi}}\bm{u}(t) = \tilde{\bm{\Pi}}\bm{R}(\bm{u}(t))\nonumber\\
	\Leftrightarrow\quad&\frac{\partial}{\partial t} \tilde{\bm{\Pi}}(\bm{u}' + \tilde{\bm{u}}^*(t) + \bar{\bm{u}}^*(t)) = \tilde{\bm{\Pi}}\bm{R}(\bm{u}' + \tilde{\bm{u}}^*(t) + \bar{\bm{u}}^*(t))\nonumber\\
	\Leftrightarrow\quad&\tilde{\bm{\Pi}}(\dot{\tilde{\bm{u}}}^*(t) + \dot{\bar{\bm{u}}}^*(t)) = \tilde{\bm{\Pi}}\bm{R}(\bm{u}' + \tilde{\bm{u}}^*(t) + \bar{\bm{u}}^*(t)).\label{eq:FOM_Projected_divided1}
\end{align}
It can be shown that if the coarse-scale operator is applied to the coarse-scale velocity field, the resulting output is the coarse-scale velocity itself, i.e.,
\begin{align}\label{eq:coarse_scale_projection_coarse_scale_velocity}
	\tilde{\bm{\Pi}}\tilde{\bm{u}}^*(t) = \tilde{\bm{\Phi}}\tilde{\bm{\Phi}}^T \tilde{\bm{\Phi}} \tilde{\bm{a}}(t) = \tilde{\bm{\Phi}} \tilde{\bm{a}}(t) = \tilde{\bm{u}}^*(t),
\end{align}
Conversely, the same operator applied to the fine-scale velocity field yields zero, i.e., 
\begin{align}\label{eq:coarse_scale_projection_fine_scale_velocity}
	\tilde{\bm{\Pi}}\bar{\bm{u}}^*(t) = \tilde{\bm{\Phi}}\tilde{\bm{\Phi}}^T \bar{\bm{\Phi}} \bar{\bm{a}}(t) = 0.
\end{align}
This outcome is attributed to the orthonormality of the basis vectors $\bm{\phi}_i^T \bm{\phi}_k = \delta_{ik}$ and thus $\tilde{\bm{\Phi}}^T \tilde{\bm{\Phi}} = \bm{I}$ and $\tilde{\bm{\Phi}}^T \bar{\bm{\Phi}} = \bm{0}$ hold.
With~\eqref{eq:coarse_scale_projection_coarse_scale_velocity} and~\eqref{eq:coarse_scale_projection_fine_scale_velocity} the system \eqref{eq:FOM_Projected_divided} becomes
\begin{align}\label{eq:FOM_Projected_divided2}
	\dot{\tilde{\bm{u}}}^*(t)  = \tilde{\bm{\Pi}}\bm{R}(\bm{u}' + \tilde{\bm{u}}^*(t) + \bar{\bm{u}}^*(t)),
\end{align}
providing a representation of the FOM expressed with the resolved coarse scales.
For standard Galerkin ROMs the influence of the unresolved fine scales is usually neglected and \eqref{eq:FOM_Projected_divided2} becomes
\begin{align*}
	\dot{\tilde{\bm{u}}}^*(t)  := \tilde{\bm{\Pi}}\bm{R}(\bm{u}' + \tilde{\bm{u}}^*(t)).
\end{align*}
\newpage

\section{Coefficients of the G-ROM for fluid flows governed by the incompressible Navier-Stokes equations}\label{ap:Matrices_G_ROM}
This section shows the coefficients for the standard G-ROM applied to fluid flows governed by the incompressible Navier-Stokes equations from Section~\ref{sec:Galerkin_Projection}. Specifically, it introduces the computational process for the coefficient matrices $\bm{Q}^{\text{G},N} \in \mathbb{R}^{N\times r^2}$, $\bm{L}^{\text{G},N} \in \mathbb{R}^{N\times r}$, and $\bm{C}^{\text{G},N} \in \mathbb{R}^{N}$, where
\begin{equation}
	\begingroup % keep the change local
	\setlength\arraycolsep{2pt}
	\begin{aligned}
		\bm{Q}^{\text{G},N} &= - (\tilde{\bm{\Phi}} \cdot \bm{\nabla})\tilde{\bm{\Phi}}\\
		&=
		-\left(\begin{matrix}
			(\tilde{\bm{\phi}}_1(x_1) \cdot \nabla)\tilde{\bm{\phi}}_1(x_1) & \cdots & (\tilde{\bm{\phi}}_1(x_1) \cdot \nabla)\tilde{\bm{\phi}}_r(x_1) & \cdots & (\tilde{\bm{\phi}}_r(x_1) \cdot \nabla)\tilde{\bm{\phi}}_r(x_1) \\
			\vdots & \ddots & \vdots & \ddots & \vdots\\
			(\tilde{\bm{\phi}}_1(x_{N_{\text{grid}}}) \cdot \nabla)\tilde{\bm{\phi}}_1(x_{N_{\text{grid}}}) & \cdots & (\tilde{\bm{\phi}}_1(x_{N_{\text{grid}}}) \cdot \nabla)\tilde{\bm{\phi}}_r(x_{N_{\text{grid}}}) & \cdots & (\tilde{\bm{\phi}}_r(x_{N_{\text{grid}}}) \cdot \nabla)\tilde{\bm{\phi}}_r(x_{N_{\text{grid}}}) \\
		\end{matrix}\right)\\
		&=
		\left(\begin{matrix}
			\bm{Q}^{\text{G},N}_{1,1}(x_1) & \cdots & \bm{Q}^{\text{G},N}_{1,r}(x_1) & \cdots &  \bm{Q}^{\text{G},N}_{r,r}(x_1)\\
			\vdots & \ddots & \vdots & \ddots & \vdots\\
			\bm{Q}^{\text{G},N}_{1,1}(x_{N_{\text{grid}}}) & \cdots & \bm{Q}^{\text{G},N}_{1,r}(x_{N_{\text{grid}}}) & \cdots & \bm{Q}^{\text{G},N}_{r,r}(x_{N_{\text{grid}}}) \\
		\end{matrix}\right),
	\end{aligned}
	\endgroup
\end{equation}\\
\begin{equation}
	\begingroup % keep the change local
	\begin{aligned}
		\bm{L}^{\text{G},N} =& -(\tilde{\bm{\Phi}} \cdot \bm{\nabla})\bm{u}' - (\bm{u}' \cdot \bm{\nabla})\tilde{\bm{\Phi}} + \nu\bm{\Delta} \tilde{\bm{\Phi}}\\
		=&
		-\left(\begin{matrix}
			(\tilde{\bm{\phi}}_1(x_1) \cdot \nabla)u'(x_1) & \cdots & \tilde{\bm{\phi}}_r(x_1) \cdot \nabla)u'(x_1) \\
			\vdots & \ddots & \vdots\\
			(\tilde{\bm{\phi}}_1(x_{N_{\text{grid}}}) \cdot \nabla)u'(x_{N_{\text{grid}}}) & \cdots & \tilde{\bm{\phi}}_r(x_{N_{\text{grid}}}) \cdot \nabla)u'(x_{N_{\text{grid}}})
		\end{matrix}\right)\\
		&-\left(\begin{matrix}
			(u'(x_1) \cdot \nabla) \tilde{\bm{\phi}}_1(x_1) & \cdots & (u'(x_1) \cdot \nabla) \tilde{\bm{\phi}}_r(x_1) \\
			\vdots & \ddots & \vdots\\
			(u'(x_{N_{\text{grid}}}) \cdot \nabla ) \tilde{\bm{\phi}}_1(x_{N_{\text{grid}}}) & \cdots & (u'(x_{N_{\text{grid}}}) \cdot \nabla) \tilde{\bm{\phi}}_r(x_{N_{\text{grid}}}) 
		\end{matrix}\right)\\
		&+\left(\begin{matrix}
			\nu\Delta \tilde{\bm{\phi}}_1(x_1) & \cdots & \nu\Delta \tilde{\bm{\phi}}_r(x_1) \\
			\vdots & \ddots & \vdots\\
			\nu\Delta \tilde{\bm{\phi}}_1(x_{N_{\text{grid}}}) & \cdots & \nu\Delta \tilde{\bm{\phi}}_r(x_{N_{\text{grid}}})
		\end{matrix}\right)\\
		=&
		\left(\begin{matrix}
			\bm{L}^{\text{G},N}_1(x_1) & \cdots & \bm{L}^{\text{G},N}_r(x_1) \\
			\vdots & \ddots & \vdots\\
			\bm{L}^{\text{G},N}_1(x_{N_{\text{grid}}}) & \cdots & \bm{L}^{\text{G},N}_r(x_{N_{\text{grid}}})
		\end{matrix}\right),
	\end{aligned}
	\endgroup
\end{equation}\\
\begin{equation}
	\begingroup % keep the change local
	\begin{aligned}
		\bm{C}^{\text{G},N} =& -(\bm{u}' \cdot \bm{\nabla})\bm{u}' +\nu\bm{\Delta} \bm{u}'\\
		=&
		-\left(\begin{matrix}
			(u'(x_1) \cdot \nabla) u'(x_1)\\
			\vdots\\
			(u'(x_{N_{\text{grid}}}) \cdot \nabla ) u'(x_{N_{\text{grid}}}) 
		\end{matrix}\right)
		+\left(\begin{matrix}
			\nu\Delta u'(x_1) \\
			\vdots\\
			\nu\Delta u'(x_{N_{\text{grid}}})
		\end{matrix}\right)\\
		=&
		\left(\begin{matrix}
			\bm{C}^{\text{G},N}(x_1)\\
			\vdots\\
			\bm{C}^{\text{G},N}(x_{N_{\text{grid}}}) 
		\end{matrix}\right).
	\end{aligned}
	\endgroup
\end{equation}

\newpage
\section{Coefficients of the eAPG-ROM for fluid flows governed by the incompressible Navier-Stokes equations}\label{ap:Matrices_APG_ROM}
This section shows the coefficients for the eAPG-ROM applied to fluid flows governed by the incompressible Navier-Stokes equations from Section~\ref{sec:APG_ROM_NSE}. Specifically, it introduces the computational process for the coefficient matrices $\bm{K}^{\text{eAPG},N} \in \mathbb{R}^{N\times r^3}$, $\bm{Q}^{\text{eAPG},N} \in \mathbb{R}^{N\times r^2}$, $\bm{L}^{\text{eAPG},N} \in \mathbb{R}^{N\times r}$, and $\bm{C}^{\text{eAPG},N} \in \mathbb{R}^{N}$, where

\begin{equation}
	\begingroup % keep the change local
	\begin{aligned}
		\bm{K}^{\text{eAPG},N} =& - \bm{\nabla} \tilde{\bm{\Phi}} \bm{Q}^{\bar{\bm{\Pi}}}
		- (\tilde{\bm{\Phi}} \cdot\bm{\nabla}) \bm{Q}^{\bar{\bm{\Pi}}}\\
		=&-
		\left(\begin{matrix}
			\nabla \tilde{\bm{\phi}}_1(x_1) \bm{Q}^{\bar{\bm{\Pi}}}_{1,1}(x_1) &
			\cdots &
			\nabla \tilde{\bm{\phi}}_1(x_1) \bm{Q}^{\bar{\bm{\Pi}}}_{1,r}(x_1) &
			\cdots \\
			\vdots & \ddots & \vdots & \ddots\\
			\nabla \tilde{\bm{\phi}}_1(x_{N_{\text{grid}}}) \bm{Q}^{\bar{\bm{\Pi}}}_{1,1}(x_{N_{\text{grid}}}) &
			\cdots &
			\nabla \tilde{\bm{\phi}}_1(x_{N_{\text{grid}}}) \bm{Q}^{\bar{\bm{\Pi}}}_{1,r}(x_{N_{\text{grid}}}) &
			\cdots\\
		\end{matrix}\right.\qquad\cdots\\[2ex]
		&\qquad\qquad\cdots\qquad\left.\begin{matrix}
			\nabla \tilde{\bm{\phi}}_1(x_1) \bm{Q}^{\bar{\bm{\Pi}}}_{r,r}(x_1) &
			\cdots &
			\nabla \tilde{\bm{\phi}}_r(x_1) \bm{Q}^{\bar{\bm{\Pi}}}_{r,r}(x_1) \\
			\vdots & \ddots & \vdots\\
			\nabla \tilde{\bm{\phi}}_1(x_{N_{\text{grid}}}) \bm{Q}^{\bar{\bm{\Pi}}}_{r,r}(x_{N_{\text{grid}}}) &
			\cdots &
			\nabla \tilde{\bm{\phi}}_r(x_{N_{\text{grid}}}) \bm{Q}^{\bar{\bm{\Pi}}}_{r,r}(x_{N_{\text{grid}}}) \\
		\end{matrix}\right)\\[2ex]
		&-
		\left(\begin{matrix}
			(\tilde{\bm{\phi}}_1(x_1) \cdot\nabla) \bm{Q}^{\bar{\bm{\Pi}}}_{1,1}(x_1) &
			\cdots &
			(\tilde{\bm{\phi}}_1(x_1) \cdot\nabla) \bm{Q}^{\bar{\bm{\Pi}}}_{1,r}(x_1) &
			\cdots \\
			\vdots & \ddots & \vdots & \ddots\\
			(\tilde{\bm{\phi}}_1(x_{N_{\text{grid}}}) \cdot\nabla) \bm{Q}^{\bar{\bm{\Pi}}}_{1,1}(x_{N_{\text{grid}}}) &
			\cdots &
			(\tilde{\bm{\phi}}_1(x_{N_{\text{grid}}}) \cdot\nabla) \bm{Q}^{\bar{\bm{\Pi}}}_{1,r}(x_{N_{\text{grid}}}) &
			\cdots\\
		\end{matrix}\right.\qquad\cdots\\[2ex]
		&\qquad\qquad\cdots\qquad\left.\begin{matrix}
			(\tilde{\bm{\phi}}_1(x_1) \cdot\nabla) \bm{Q}^{\bar{\bm{\Pi}}}_{r,r}(x_1) &
			\cdots &
			(\tilde{\bm{\phi}}_r(x_1) \cdot\nabla) \bm{Q}^{\bar{\bm{\Pi}}}_{r,r}(x_1) \\
			\vdots & \ddots & \vdots\\
			(\tilde{\bm{\phi}}_1(x_{N_{\text{grid}}}) \cdot\nabla) \bm{Q}^{\bar{\bm{\Pi}}}_{r,r}(x_{N_{\text{grid}}}) &
			\cdots &
			(\tilde{\bm{\phi}}_r(x_{N_{\text{grid}}}) \cdot\nabla) \bm{Q}^{\bar{\bm{\Pi}}}_{r,r}(x_{N_{\text{grid}}}) \\
		\end{matrix}\right),
	\end{aligned}
	\endgroup
\end{equation}\\
\begin{equation}
	\begingroup % keep the change local
	\setlength\arraycolsep{2pt}
	\begin{aligned}
		\bm{Q}^{\text{eAPG},N} =& 	- \bm{\nabla} \tilde{\bm{\Phi}} \bm{L}^{\bar{\bm{\Pi}}}
									- (\tilde{\bm{\Phi}} \cdot\bm{\nabla}) \bm{L}^{\bar{\bm{\Pi}}}
									- \bm{\nabla} \bm{u}' \bm{Q}^{\bar{\bm{\Pi}}}
									- (\bm{u}'\cdot\bm{\nabla}) \bm{Q}^{\bar{\bm{\Pi}}}
									+ \nu\bm{\Delta} \bm{Q}^{\bar{\bm{\Pi}}}\\
		=&-
		\left(\begin{matrix}
			\nabla \tilde{\bm{\phi}}_1(x_1) \bm{L}^{\bar{\bm{\Pi}}}_{1}(x_1) &
			\cdots &
			\nabla \tilde{\bm{\phi}}_1(x_1) \bm{L}^{\bar{\bm{\Pi}}}_{r}(x_1) &
			\cdots &
			\nabla \tilde{\bm{\phi}}_r(x_1) \bm{L}^{\bar{\bm{\Pi}}}_{r}(x_1) \\
			\vdots & \ddots & \vdots & \ddots & \vdots \\
			\nabla \tilde{\bm{\phi}}_1(x_{N_{\text{grid}}}) \bm{L}^{\bar{\bm{\Pi}}}_{1}(x_{N_{\text{grid}}}) &
			\cdots &
			\nabla \tilde{\bm{\phi}}_1(x_{N_{\text{grid}}}) \bm{L}^{\bar{\bm{\Pi}}}_{r}(x_{N_{\text{grid}}}) &
			\cdots &
			\nabla \tilde{\bm{\phi}}_r(x_{N_{\text{grid}}}) \bm{L}^{\bar{\bm{\Pi}}}_{r}(x_{N_{\text{grid}}}) \\
		\end{matrix}\right)\\[2ex]
		&-
		\left(\begin{matrix}
			(\tilde{\bm{\phi}}_1(x_1) \cdot\nabla) \bm{L}^{\bar{\bm{\Pi}}}_{1}(x_1) &
			\cdots &
			(\tilde{\bm{\phi}}_1(x_1) \cdot\nabla) \bm{L}^{\bar{\bm{\Pi}}}_{r}(x_1) &
			\cdots &
			(\tilde{\bm{\phi}}_r(x_1) \cdot\nabla) \bm{L}^{\bar{\bm{\Pi}}}_{r}(x_1) \\
			\vdots & \ddots & \vdots & \ddots & \vdots \\
			(\tilde{\bm{\phi}}_1(x_{N_{\text{grid}}}) \cdot\nabla) \bm{L}^{\bar{\bm{\Pi}}}_{1}(x_{N_{\text{grid}}}) &
			\cdots &
			(\tilde{\bm{\phi}}_1(x_{N_{\text{grid}}}) \cdot\nabla) \bm{L}^{\bar{\bm{\Pi}}}_{r}(x_{N_{\text{grid}}}) &
			\cdots &
			(\tilde{\bm{\phi}}_r(x_{N_{\text{grid}}}) \cdot\nabla) \bm{L}^{\bar{\bm{\Pi}}}_{r}(x_{N_{\text{grid}}}) \\
		\end{matrix}\right)\\[2ex]
		&-
		\left(\begin{matrix}
			\nabla u'(x_1) \bm{Q}^{\bar{\bm{\Pi}}}_{1,1}(x_1) &
			\cdots &
			\nabla u'(x_1) \bm{Q}^{\bar{\bm{\Pi}}}_{1,r}(x_1) &
			\cdots &
			\nabla u'(x_1) \bm{Q}^{\bar{\bm{\Pi}}}_{r,r}(x_1) \\
			\vdots & \ddots & \vdots & \ddots & \vdots \\
			\nabla u'(x_{N_{\text{grid}}}) \bm{Q}^{\bar{\bm{\Pi}}}_{1,1}(x_{N_{\text{grid}}}) &
			\cdots &
			\nabla u'(x_{N_{\text{grid}}}) \bm{Q}^{\bar{\bm{\Pi}}}_{1,r}(x_{N_{\text{grid}}}) &
			\cdots &
			\nabla u'(x_{N_{\text{grid}}}) \bm{Q}^{\bar{\bm{\Pi}}}_{r,r}(x_{N_{\text{grid}}}) \\
		\end{matrix}\right)\\[2ex]
		&-
		\left(\begin{matrix}
			(u'(x_1) \cdot\nabla) \bm{Q}^{\bar{\bm{\Pi}}}_{1,1}(x_1) &
			\cdots &
			(u'(x_1) \cdot\nabla) \bm{Q}^{\bar{\bm{\Pi}}}_{1,r}(x_1) &
			\cdots &
			(u'(x_1) \cdot\nabla) \bm{Q}^{\bar{\bm{\Pi}}}_{r,r}(x_1) \\
			\vdots & \ddots & \vdots & \ddots & \vdots \\
			(u'(x_{N_{\text{grid}}}) \cdot\nabla) \bm{Q}^{\bar{\bm{\Pi}}}_{1,1}(x_{N_{\text{grid}}}) &
			\cdots &
			(u'(x_{N_{\text{grid}}}) \cdot\nabla) \bm{Q}^{\bar{\bm{\Pi}}}_{1,r}(x_{N_{\text{grid}}}) &
			\cdots &
			(u'(x_{N_{\text{grid}}}) \cdot\nabla) \bm{Q}^{\bar{\bm{\Pi}}}_{r,r}(x_{N_{\text{grid}}}) \\
		\end{matrix}\right)\\[2ex]
		&+
		\left(\begin{matrix}
			\nu\Delta \bm{Q}^{\bar{\bm{\Pi}}}_{1,1}(x_1) &
			\cdots &
			\nu\Delta \bm{Q}^{\bar{\bm{\Pi}}}_{1,r}(x_1)  &
			\cdots &
			\nu\Delta \bm{Q}^{\bar{\bm{\Pi}}}_{r,r}(x_1)  \\
			\vdots & \ddots & \vdots & \ddots & \vdots \\
			\nu\Delta \bm{Q}^{\bar{\bm{\Pi}}}_{1,1}(x_{N_{\text{grid}}})  &
			\cdots &
			\nu\Delta \bm{Q}^{\bar{\bm{\Pi}}}_{1,r}(x_{N_{\text{grid}}}) &
			\cdots &
			\nu\Delta \bm{Q}^{\bar{\bm{\Pi}}}_{r,r}(x_{N_{\text{grid}}})
		\end{matrix}\right),
	\end{aligned}
	\endgroup
\end{equation}\\
\begin{equation}
	\begingroup % keep the change local
	\begin{aligned}
		\bm{L}^{\text{eAPG},N} =& 	- \bm{\nabla} \tilde{\bm{\Phi}} \bm{C}^{\bar{\bm{\Pi}}}
									- (\tilde{\bm{\Phi}} \cdot\bm{\nabla}) \bm{C}^{\bar{\bm{\Pi}}}
									- \bm{\nabla} \bm{u}' \bm{L}^{\bar{\bm{\Pi}}}
									- (\bm{u}'\cdot\bm{\nabla}) \bm{L}^{\bar{\bm{\Pi}}}
									+ \nu\bm{\Delta} \bm{L}^{\bar{\bm{\Pi}}}\\
		=&-
		\left(\begin{matrix}
			\nabla \tilde{\bm{\phi}}_1(x_1) \bm{C}^{\bar{\bm{\Pi}}}(x_1) &
			\cdots &
			\nabla \tilde{\bm{\phi}}_1(x_1) \bm{C}^{\bar{\bm{\Pi}}}(x_1) \\
			\vdots & \ddots & \vdots  \\
			\nabla \tilde{\bm{\phi}}_1(x_{N_{\text{grid}}}) \bm{C}^{\bar{\bm{\Pi}}}(x_{N_{\text{grid}}}) &
			\cdots &
			\nabla \tilde{\bm{\phi}}_1(x_{N_{\text{grid}}}) \bm{C}^{\bar{\bm{\Pi}}}(x_{N_{\text{grid}}}) \
		\end{matrix}\right)\\[2ex]
		&-
		\left(\begin{matrix}
			(\tilde{\bm{\phi}}_1(x_1) \cdot\nabla) \bm{C}^{\bar{\bm{\Pi}}}(x_1) &
			\cdots &
			(\tilde{\bm{\phi}}_1(x_1) \cdot\nabla) \bm{C}^{\bar{\bm{\Pi}}}(x_1) \\
			\vdots & \ddots & \vdots \\
			(\tilde{\bm{\phi}}_1(x_{N_{\text{grid}}}) \cdot\nabla) \bm{C}^{\bar{\bm{\Pi}}}(x_{N_{\text{grid}}}) &
			\cdots &
			(\tilde{\bm{\phi}}_1(x_{N_{\text{grid}}}) \cdot\nabla) \bm{C}^{\bar{\bm{\Pi}}}(x_{N_{\text{grid}}}) \\
		\end{matrix}\right)\\[2ex]
		&-
		\left(\begin{matrix}
			\nabla u'(x_1) \bm{L}^{\bar{\bm{\Pi}}}_{1}(x_1) &
			\cdots &
			\nabla u'(x_1) \bm{L}^{\bar{\bm{\Pi}}}_{r}(x_1) \\
			\vdots & \ddots & \vdots \\
			\nabla u'(x_{N_{\text{grid}}}) \bm{L}^{\bar{\bm{\Pi}}}_{1}(x_{N_{\text{grid}}}) &
			\cdots &
			\nabla u'(x_{N_{\text{grid}}}) \bm{L}^{\bar{\bm{\Pi}}}_{r}(x_{N_{\text{grid}}}) \\
		\end{matrix}\right)\\[2ex]
		&-
		\left(\begin{matrix}
			(u'(x_1) \cdot\nabla) \bm{L}^{\bar{\bm{\Pi}}}_{1}(x_1) &
			\cdots &
			(u'(x_1) \cdot\nabla) \bm{L}^{\bar{\bm{\Pi}}}_{r}(x_1) \\
			\vdots & \ddots & \vdots \\
			(u'(x_{N_{\text{grid}}}) \cdot\nabla) \bm{L}^{\bar{\bm{\Pi}}}_{1}(x_{N_{\text{grid}}}) &
			\cdots &
			(u'(x_{N_{\text{grid}}}) \cdot\nabla) \bm{L}^{\bar{\bm{\Pi}}}_{r}(x_{N_{\text{grid}}}) \\
		\end{matrix}\right)\\[2ex]
		&+
		\left(\begin{matrix}
			\nu\Delta \bm{L}^{\bar{\bm{\Pi}}}_{1}(x_1) &
			\cdots &
			\nu\Delta \bm{L}^{\bar{\bm{\Pi}}}_{r}(x_1) \\
			\vdots & \ddots & \vdots \\
			\nu\Delta \bm{L}^{\bar{\bm{\Pi}}}_{1}(x_{N_{\text{grid}}})  &
			\cdots &
			\nu\Delta \bm{L}^{\bar{\bm{\Pi}}}_{r}(x_{N_{\text{grid}}})
		\end{matrix}\right),
	\end{aligned}
	\endgroup
\end{equation}\\
\begin{equation}
	\begingroup % keep the change local
	\begin{aligned}
		\bm{C}^{\text{eAPG},N} =& 	- \bm{\nabla} \bm{u}' \bm{C}^{\bar{\bm{\Pi}}}
									- (\bm{u}'\cdot\bm{\nabla}) \bm{C}^{\bar{\bm{\Pi}}}
									+ \nu\bm{\Delta} \bm{C}^{\bar{\bm{\Pi}}}\\
		=&-
		\left(\begin{matrix}
			\nabla u'(x_1) \bm{C}^{\bar{\bm{\Pi}}}(x_1)  \\
			\vdots \\
			\nabla u'(x_{N_{\text{grid}}}) \bm{C}^{\bar{\bm{\Pi}}}(x_{N_{\text{grid}}}) \\
		\end{matrix}\right)
		-
		\left(\begin{matrix}
			(u'(x_1) \cdot\nabla) \bm{C}^{\bar{\bm{\Pi}}}(x_1) \\
			\vdots \\
			(u'(x_{N_{\text{grid}}}) \cdot\nabla) \bm{C}^{\bar{\bm{\Pi}}}(x_{N_{\text{grid}}}) \\
		\end{matrix}\right)
		+
		\left(\begin{matrix}
			\nu\Delta \bm{C}^{\bar{\bm{\Pi}}}(x_1) \\
			\vdots \\
			\nu\Delta \bm{C}^{\bar{\bm{\Pi}}}(x_{N_{\text{grid}}})  
		\end{matrix}\right).
	\end{aligned}
	\endgroup
\end{equation}

\typeout{col width is \the\textwidth} 
%textwidth in cm: \printinunitsof{cm}\prntlen{\textwidth}
\end{document}